\newif\ifdraft
\draftfalse

\newif\ifacm
\acmtrue

\newif\ifsubmit
\submittrue

\newif\ifanonymized
\ifdraft
  \anonymizedfalse  %
\else
\fi

\ifdraft
   \acmfalse
   \documentclass[onecolumn]{article}
   \usepackage[text={6.5in,9in},centering]{geometry}	  %
   \usepackage{setspace}
   \usepackage{times}
\else
    \ifacm

        \documentclass[acmlarge]{acmart} %
    \else
        \documentclass{sig-alternate}
    \fi
\fi

\ifsubmit
    \relax  %
\else
\fi
\ifdraft
  \usepackage{fancyhdr}	%
\else
  \relax  %
\fi
\usepackage[labelfont={sf,bf}, textfont=sf]{caption,subfig}
\usepackage[nolist]{acronym}	%
\usepackage{booktabs}   %
\usepackage{caption}
\usepackage{amsmath}
\usepackage{hyperref}

\usepackage{color}
\usepackage[mmddyyyy,hhmmss]{datetime}
\usepackage{microtype}	%
\usepackage{siunitx}		%
\usepackage{url}
\usepackage{pgfplots}
\pgfplotsset{compat=1.15}
\usepackage{xspace}	%
\usepackage{graphicx}
\usepackage{wrapfig}
\usepackage{lipsum} %
\DeclareGraphicsExtensions{.pdf,.png,.eps,.ps,.jpg}
\usepackage{arydshln}
\usepackage{multirow}
\ifdraft  %
 \hypersetup{              %
   pdfcreator={},          %
   bookmarksnumbered=false,
   colorlinks=true,        %
   citecolor=black,        %
   filecolor=black,        %
   linkcolor=black,        %
   urlcolor=blue,         %
   bookmarksopen=false,
   bookmarks=false,
   pdfpagemode=UseNone,   %
   breaklinks=true,
   pdfstartview={FitH},
 }
\fi

\newcommand{\hide}[1]{\relax}
\ifsubmit

  \newcommand{\colorcomment}[3]{\relax}
  \newcommand{\greencomment}[2]{\relax}
  \newcommand{\yellowcomment}[2]{\relax}
  \newcommand{\cyancomment}[2]{\relax}
  \newcommand{\bluecomment}[2]{\relax}
  \newcommand{\redcomment}[2]{\relax}
  \newcommand{\magentacomment}[2]{\relax}

  \newcommand{\hey}[1]{\relax}
  \newcommand{\bibtex}[1]{\relax}
  \newcommand{\task}[1]{\relax}
  \newcommand{\question}[1]{\relax}
  \newcommand{\note}[1]{\relax}

\else
  \newcommand{\colorcomment}[3]{\par\noindent\textcolor{#1}{[{#2}: {#3}]}\par} %
  \newcommand{\greencomment}[2]{\colorcomment{green}{#1}{#2}} %
  \newcommand{\yellowcomment}[2]{\colorcomment{yellow}{#1}{#2}} %
  \newcommand{\cyancomment}[2]{\colorcomment{cyan}{#1}{#2}} %
  \newcommand{\bluecomment}[2]{\colorcomment{blue}{#1}{#2}}
  \newcommand{\redcomment}[2]{\colorcomment{red}{#1}{#2}}
  \newcommand{\magentacomment}[2]{\colorcomment{magenta}{#1}{#2}}

  \newcommand{\hey}[1]{\textcolor{blue}{[{#1}]}} %
  \newcommand{\note}[2]{\magentacomment{Note}{#1}} %
  \newcommand{\bibtex}[1]{\textcolor{magenta}{@bibtex}\{#1\}} %
  \newcommand{\task}[1]{\redcomment{Task}{#1}} %
  \newcommand{\question}[1]{\bluecomment{Q?}{#1}} %

\fi

\definecolor{red}{rgb}{0,0,0}

\ifdraft

\else

    \ifacm
      \setcopyright{rightsretained} 
        \acmJournal{IMWUT}
        \acmYear{2021} \acmVolume{5} \acmNumber{2} \acmArticle{74} \acmMonth{6} \acmPrice{}\acmDOI{10.1145/3463492}

        
        
    \else
    \fi
        
\fi
\def\papertitle{Detecting Receptivity for mHealth Interventions in the Natural Environment}  %
\def\papershorttitle{Detecting Receptivity for mHealth Interventions in the Natural Environment}  %
\begin{document}

\newcommand{\appname}{Ally+}

\ifacm
    \title[\papershorttitle]{\papertitle}
\else
	\title{\papertitle}
\fi

\ifanonymized %
	\author{Anonymized for blind submission}
\else
    \ifacm
        \author{Varun Mishra}
        \authornote{Both authors contributed equally to this research.}
        \authornote{Corresponding author.}
        \email{varun@cs.dartmouth.edu}
        \affiliation{%
          \institution{Dartmouth College}
        }
        
        \author{Florian K\"{u}nzler}
        \authornotemark[1]
        \affiliation{%
          \institution{ETH Z\"{u}rich}
        }
        
        \author{Jan-Niklas Kramer}
        \affiliation{%
          \institution{University of St. Gallen}
        }
        
        \author{Elgar Fleisch}
        \affiliation{%
          \institution{ETH Z\"{u}rich}
        }
          \affiliation{%
          \institution{University of St. Gallen}
        }

        \author{Tobias Kowatsch}
        \affiliation{%
          \institution{ETH Z\"{u}rich}
        }
          \affiliation{%
          \institution{University of St. Gallen}
        }
           \affiliation{%
          \institution{National University of Singapore}
        }
        
        \author{David Kotz}
        \affiliation{%
          \institution{Dartmouth College}
        }
    \else
        	\author{Varun Mishra$^{1,*}$, Florian K\"{u}nzler$^{2,*}$, Jan-Niklas Kramer$^3$, Elgar Fleisch$^2$,\\ Tobias Kowatsch$^3$ and David Kotz$^1$\\  %
	\small $^1$Dartmouth College, $^2$ETH Z\"{u}rich, $^3$University of St. Gallen\\
	\small $^*$Equal Contributions
	}
\fi
\renewcommand{\shortauthors}{Mishra and K\"{u}nzler, et al.}

\date{}

\begin{acronym}
\acro{LOGO}{Leave-One-Group-Out}
\acro{IRB}{Institutional Review Board}
\acro{LR}{Logistic Regression}
\acro{JIT}{Just-In-Time}
\acro{JITI}{Just-In-Time Intervention}
\acro{JITAI}{Just-In-Time Adaptive Intervention}
\acro{EMA}{Ecological Momentary Assessment}
\acro{GLM}{Generalized Linear Model}
\acro{ANOVA}{Analysis of Variance}
\acro{ML}{machine learning}
\end{acronym}

\ifdraft
    \maketitle
\begin{abstract}
\ac{JITAI} is an emerging technique with great potential to support health behavior by providing the right type and amount of support at the right time. A crucial aspect of \ac{JITAI}s is properly timing the delivery of interventions, to ensure that a user is receptive and ready to process and use the support provided. Some prior works have explored the association of context and some user-specific traits on receptivity, and have built post-study machine-learning models to detect receptivity. For effective intervention delivery, however, a \ac{JITAI} system needs to make in-the-moment decisions about a user's receptivity. To this end, we conducted a study in which we deployed machine-learning models to detect receptivity in the natural environment, i.e., in free-living conditions. 

We leveraged prior work regarding receptivity to \ac{JITAI}s and deployed a chatbot-based digital coach~-- Ally~-- that provided physical-activity interventions and motivated participants to achieve their step goals. We extended the original Ally~app to include two types of machine-learning model that used contextual information about a person to predict when a person is receptive: a \textit{static model\/} that was built before the study started and remained constant for all participants and an \textit{adaptive model\/} that continuously learned the receptivity of individual participants and updated itself as the study progressed. For comparison, we included a \textit{control model\/} that sent intervention messages at random times. The app randomly selected a delivery model for each intervention message. We observed that the machine-learning models led up to a 40\% improvement in receptivity as compared to the control model. Further, we evaluated the temporal dynamics of the different models and observed that receptivity to messages from the adaptive model increased over the course of the study.

\end{abstract}
 \else
    \ifacm
\begin{abstract}
\ac{JITAI} is an emerging technique with great potential to support health behavior by providing the right type and amount of support at the right time. A crucial aspect of \ac{JITAI}s is properly timing the delivery of interventions, to ensure that a user is receptive and ready to process and use the support provided. Some prior works have explored the association of context and some user-specific traits on receptivity, and have built post-study machine-learning models to detect receptivity. For effective intervention delivery, however, a \ac{JITAI} system needs to make in-the-moment decisions about a user's receptivity. To this end, we conducted a study in which we deployed machine-learning models to detect receptivity in the natural environment, i.e., in free-living conditions. 

We leveraged prior work regarding receptivity to \ac{JITAI}s and deployed a chatbot-based digital coach~-- Ally~-- that provided physical-activity interventions and motivated participants to achieve their step goals. We extended the original Ally~app to include two types of machine-learning model that used contextual information about a person to predict when a person is receptive: a \textit{static model\/} that was built before the study started and remained constant for all participants and an \textit{adaptive model\/} that continuously learned the receptivity of individual participants and updated itself as the study progressed. For comparison, we included a \textit{control model\/} that sent intervention messages at random times. The app randomly selected a delivery model for each intervention message. We observed that the machine-learning models led up to a 40\% improvement in receptivity as compared to the control model. Further, we evaluated the temporal dynamics of the different models and observed that receptivity to messages from the adaptive model increased over the course of the study.

\end{abstract}
         \maketitle  %
    \else
        \maketitle
\begin{abstract}
\ac{JITAI} is an emerging technique with great potential to support health behavior by providing the right type and amount of support at the right time. A crucial aspect of \ac{JITAI}s is properly timing the delivery of interventions, to ensure that a user is receptive and ready to process and use the support provided. Some prior works have explored the association of context and some user-specific traits on receptivity, and have built post-study machine-learning models to detect receptivity. For effective intervention delivery, however, a \ac{JITAI} system needs to make in-the-moment decisions about a user's receptivity. To this end, we conducted a study in which we deployed machine-learning models to detect receptivity in the natural environment, i.e., in free-living conditions. 

We leveraged prior work regarding receptivity to \ac{JITAI}s and deployed a chatbot-based digital coach~-- Ally~-- that provided physical-activity interventions and motivated participants to achieve their step goals. We extended the original Ally~app to include two types of machine-learning model that used contextual information about a person to predict when a person is receptive: a \textit{static model\/} that was built before the study started and remained constant for all participants and an \textit{adaptive model\/} that continuously learned the receptivity of individual participants and updated itself as the study progressed. For comparison, we included a \textit{control model\/} that sent intervention messages at random times. The app randomly selected a delivery model for each intervention message. We observed that the machine-learning models led up to a 40\% improvement in receptivity as compared to the control model. Further, we evaluated the temporal dynamics of the different models and observed that receptivity to messages from the adaptive model increased over the course of the study.

\end{abstract}
     \fi
\fi

\section{Introduction} %
\label{sec:introduction}

The ubiquituous presence of mobile technologies has enabled a wide array of research into mobile health (mHealth), from sensing health conditions to providing behavior-change interventions. In the past, ubiquitous technologies like smartphones and wearables have shown promise in detecting stress, anxiety, mood, depression, personality change, addictive behavior, physical activity and a host of other conditions~\cite{hovsepian:cstress,mishra:stress-ml,wang:studentlife14,wang:crosscheck16}. 
Furthermore, several studies have demonstrated the potential of smartphone-based digital interventions to affect positive behavior change for a range of  conditions like smoking, alcohol disorder, eating disorders, and physical inactivity~\cite{shih2019breeze, ghorai2014mhealth, kazemi2017systematic, martin2015mactive}. The eventual goal in mHealth is to be able to combine the two components of accurate sensing and effective interventions to improve the quality of life amongst people suffering from various conditions. 

\ac{JITAI} is a novel intervention design that aims to deliver the right type and amount of support, at the right time, while adapting as-needed to the users' internal and external contextual change~\cite{nahum2016just, nahum2015building}. Several studies have employed \ac{JITAI}-like interventions to affect behavior change for physical inactivity~\cite{kramer2018investigating,consolvo2008activity}, alcohol use~\cite{gustafson2014smartphone}, mental illness~\cite{ben2014feasibility}, smoking~\cite{riley2008internet}, and obesity~\cite{bauer2010enhancement}. For \ac{JITAI}s to be effective they must deliver the intervention at ``the right time,'' {\color{red}one aspect of the ``right time'' is} when a person enters a state of \emph{vulnerability}, i.e., a period of heightened susceptibility for a negative health outcome~\cite{nahum2016just}. For example, an intervention to alleviate stress levels should be delivered when a person is stressed or about to be stressed; delivering it when a person is not actually stressed would not be useful. Furthermore, it is also important to deliver interventions at a time when the participant is in a state of \emph{receptivity}, i.e., a period when the participant is able to receive, process, and use the intervention provided~\cite{nahum2016just}. For example, an intervention targeted at reducing sedentary behavior would be effective when a person is ``available'' to act on it; delivering it when a person is driving is not only sub-optimal but may also be dangerous.

Although prior mHealth research focuses on detecting states of vulnerability or on developing effective intervention mechanisms, little research has been done in identifying \emph{states-of-receptivity}. In our prior work, we developed the original Ally app to deliver physical-activity interventions and explored how the  passively collected contextual factors associated with receptivity in a study with 189 participants~\cite{kunzler:receptivity2019}. In the same prior work, we also evaluated the feasibility of building machine-learning models to detect receptivity and achieved a 77\% improvement in F1-score over a biased random model. Choi~et~al.\  conducted a 3-week study with 31 participants in which the authors collected self-reports about the participants' context and cognitive/physical state to understand association with the response to relevant \ac{JIT} support targeted towards sedentary behavior~\cite{choi:receptivity19}. Koch~et~al.\  observed how drivers interact with and respond to affective well-being interventions while driving in their daily lives, with 10 participants over two months~\cite{koch:car-receptivity}. Sarker~et~al.\  conducted a study with 30 participants to explore discriminative features and built machine-learning models to detect receptivity to \ac{JITI} with a reported 77.9\% accuracy. The authors, however, did not deliver interventions; instead they used \ac{EMA} and claimed that interactions with \ac{EMA} prompts would be similar to interactions with intervention prompts.

All of those prior works, however, focused on data collection followed by post-study analysis and evaluation of post-study machine-learning models, with the expectation that the models would perform similarly when deployed in real-life conditions. In this paper, we go beyond post-study analysis: we deployed two different machine-learning models to predict in-the-moment receptivity, and used that prediction to decide when to deliver the intervention. We deployed these models in a physical-activity app used by 83 participants in free-living conditions over a period of 3~weeks. Our goal was to evaluate whether such models actually helped increase receptivity to interventions. 

Given the promising results from our previous study, we decided to build upon that work~\cite{kunzler:receptivity2019}. 
We used the data from the previous study to build two different machine-learning models, which we later deployed in our field study: (a)~a \emph{static\/} model that remained constant for all participants through out the study, and (b)~an \emph{adaptive\/} model that continuously learnt the receptivity of individual participants from their enrollment in the study and updated the model as the study progressed; we delayed activation of this model until the participant had been in the study for 7~days, however, to ensure that enough data was collected for that participant before using the model's predictions. To compare the utility of these models, we also included (c)~a \emph{control\/} model that would send the intervention messages at a random time. 
We extended the original Ally app~\cite{kunzler:receptivity2019}, to incorporate the different models and enable \emph{in-the-moment\/} detection of receptivity.\footnote{During the first 7 days, the app randomly chose between the \emph{control\/} and \emph{static\/} models. After 7 days, the app randomly chose between the \emph{control}, \emph{static}, and \emph{adaptive\/} models.} We also translated all of their intervention messages from German to English as our new study was conducted in an English-speaking population.

\newcommand{\RQone}{On a population level, i.e., across intervention messages and across all users, does delivering interventions at a \ac{ML}-detected time lead to higher receptivity than delivering interventions at a random time?}
\newcommand{\RQtwo}{How does receptivity of individual users change when interventions are delivered  through \ac{ML}-based intervention timing vs. at random times?}
\newcommand{\RQthree}{How do the different models for predicting receptivity perform over time?}

\medskip\noindent
We explore three core research questions (RQs):
\begin{itemize}
    \item \textit{RQ-1: \RQone}
    
    For all the messages sent in the study (across all participants), we evaluated how receptivity differed between messages delivered at `opportune' moments, i.e., using the static and adaptive models, as compared to random times. 

    \item \textit{RQ-2: \RQtwo}
    
    For individual participants who received intervention messages at both (a)~model-based times and (b)~at random times, we evaluated how their individual receptivity changed with the different models.

    \item \textit{RQ-3: \RQthree}

    We evaluated how the participants' daily receptivity changed as the study progressed.  Our goal was to understand whether the models performed consistently throughout the study or whether their effectiveness improved (or worsened) as the study progressed.
    
\end{itemize}

\smallskip\noindent
While exploring the three research questions, we make the following \textbf{contributions}:

\begin{itemize}
    \item We built two separate models: (a)~static and (b)~adaptive, using previously collected data to model receptivity to chat-based \ac{JITAI}s. We deployed both the models in a 3-week long field study with 83 participants, and actually used the output of the model predictions to deliver \ac{JITAI}s \emph{in-the-moment}. Our work moves beyond post-study evaluations, and tests the effectiveness of deploying receptivity-detection models trained using data from a previous study.
    Further, this is the first work to \emph{deploy\/} an adaptive model to detect receptivity to JITAI and observe how the model performance changes as the study progresses.    

    \item We evaluated the performance of the two models with a \emph{control\/} model (which sent intervention messages at random times), both on a population level and a participant level. We observed that the \emph{static\/} model led to significantly higher receptivity than the \emph{control\/} model in most instances, for both population level and participant level analyses, suggesting that it is possible to use machine-learning models to predict in-the-moment receptivity. We observed that a participant was more likely to respond to a message delivered at a predicted \emph{opportune\/} moment than at a random moment.
    
    \item We evaluated the temporal dynamics of the different models to understand how they performed over the course of the study. We observed that while receptivity from the random model decreased over time, the static model was able to maintain consistent receptivity over time, and receptivity to messages delivered by the adaptive model improved as participants progressed in the study.
\end{itemize}

We begin with a review of related work in receptivity towards interventions, as well as closely related concepts of \emph{interruptibility\/} and \emph{engagement}. We then provide a background of the Ally app and operationalization of the different receptivity metrics, followed by our study design methodology and the results for our research questions. Finally, we conclude with a discussion of the implications, future work, and limitations of our work.

\section{Related work} %
\label{sec:related_work}

In the domain of ubiquitous computing, researchers have extensively explored a related concept of \emph{interruptibility\/}. In the context of smartphone notifications, interruptibility is defined as a person's ability to be interrupted by an incoming notification by taking an action to open or view the notification content~\cite{mehrotra2015designing}. 
This stream of research has focused on the use of push notifications to attract users' attention, while making the recipients feel less `interrupted' by the notifications. In this section, we discuss the prior research on interruptibility to smartphone notifications. We also discuss prior work in capturing participant \emph{receptivity\/} in mHealth research.

Most prior `interruptibility' research has focused on analyzing and understanding user interruptibility and on the association between various contextual factors and interruptibility. 
Researchers have studied factors such as time of day and day of week~\cite{pejovic2014interruptme, pielot2015attention, avrahami2006responsiveness, mashhadi2014myth}, location~\cite{sarker2014assessing, pielot2017beyond, mehrotra2015designing}, Bluetooth information (as a proxy for social context)~\cite{pejovic2014interruptme}, call and SMS logs (another proxy for social context)~\cite{pielot2015attention, fischer2011investigating}, Wi-Fi connectivity~\cite{pejovic2014interruptme, pielot2015attention}, and phone battery information~\cite{pielot2017beyond}. While most of these factors were found to be a predictor of users' interruptibility, some studies have also contradicted those findings. 
For example, some have found time to be a significant predictor~\cite{pejovic2014interruptme, pielot2015attention}, while others have found the opposite~\cite{mashhadi2014myth, westermann2016smartphone}. The same applies for location, where Sarker~et~al.\  and others~\cite{pielot2015attention, pejovic2014interruptme, sarker2014assessing} found location was an important indicator of interruptibility and Mehrotra~et~al.\  found otherwise~\cite{mehrotra2015designing}.

Other research investigated personality traits and mental state as potential predictors for receptivity. Happy and energetic participants showed a higher availability to interruptions, compared to stressed study participants~\cite{sarker2014assessing}. Other researchers have shown the significance of personality traits to predict the response delay; in particular, neuroticism and extroversion were found to be significant~\cite{mehrotra2016my}.

Many studies have found a significant correlation between physical activity and interruptibility~\cite{okoshi2015reducing, sarker2014assessing, ho2005using, mehrotra2015designing}. The type of physical activity was also found to be significant; for example, people driving in a vehicle replied more slowly than people walking outside~\cite{sarker2014assessing}. Further, `breakpoints' in physical activity, e.g., from walking to standing, were found to be favorable times to trigger notifications~\cite{okoshi2015reducing}. Generalizing results, however, is difficult as small sample size and homogeneity of study participants is a common problem in this research area~\cite{kunzler2017efficacy}.

A few researchers have explored the concept of \emph{engagement\/}. In the context of smartphone notifications, engagement usually follows after a person is interrupted and refers to the involvement of an user in a task or app that attracts and holds the user's attention~\cite{pielot2017beyond}. Pielot~et~al.\  conducted a study ($n>330$) in which they delivered eight different types of content and observed participants' engagement and responsiveness~\cite{pielot2017beyond}. They then built predictive models to classify whether a participant would engage with the notification content, and found that their models led to an improvement of more than 66\% over a baseline classifier. Related to engagement, Dingler~et~al.\  built an app aimed at improving users' foreign-language vocabulary through notifications and app usage through out the day~\cite{dingler:learning-on-the-go}. The authors found that several contextual factors relating to phone usage, e.g., number of phone unlocks in the last 5 minutes, time since last unlock, and number of notifications in the last 5 minutes, showed significant correlations to predict users' engagement with the content.

Some works have even deployed a machine-learning classifier to infer interruptibility.
Okoshi~et~al.\  deployed a breakpoint-detection system to time notifications~\cite{okoshi:yahoo17}.
Based on a study with over 680,000 users, they found that response time to notifications was up to 49\% lower if they were delivered during activity breakpoints.
Pielot~et~al.\  deployed a model that allowed them to deliver entertaining content when the participant was likely bored; at such times, they found participants were more likely to engage~\cite{pielot2015attention}.

In the domain of \ac{JITAI}s and mHealth interventions, \emph{receptivity\/} is defined as the person's ability to receive, process, and use the support (intervention) provided~\cite{nahum2016just}. To compare with interruptibility and engagement, as highlighted by K\"{u}nzler~et~al., ``receptivity may be (loosely) conceptualized to encompass the combination of interruptibility (willingness to receive an intervention), engagement (receive the intervention), and the person's subjective perception of the intervention provided (process
and use the intervention)''~\cite{kunzler:receptivity2019}.

There have been a growing number of works exploring receptivity and interruptibility in the domain of mHealth.
Sarker~et~al.\  conducted a study with 30 participants, and identified various contextual and physiological features for their machine-learning model to detect receptivity to \ac{EMA} prompts~\cite{sarker2014assessing}. The authors drew a parallel between \ac{EMA} and interventions by claiming that interaction with self-report or \ac{EMA} prompts and interaction with interventions would be similar. The authors also provided monetary incentives for \ac{EMA} completion, however, which may have influenced the participants' receptivity. Further, Mishra~et~al.\  investigated contextual breakpoints and how they could be used to detect receptivity to \ac{EMA}~\cite{mishra:ema-workshop}.

Choi~et~al.\  conducted a 3-week study with 31 participants in which the participants reported information about their  context and cognitive/physical state; the paper explores the association of context and cognitive state with the relevant \ac{JIT} support targeted towards sedentary behavior~\cite{choi:receptivity19}. The authors identified several key factors relating to receptivity and showed that receptivity to \ac{JIT} interventions is nuanced and context-dependant.

In our prior work, we developed a smartphone app to deliver physical-activity interventions and deployed it in a study with 189 participants~\cite{kunzler:receptivity2019}. We explored how passively collected contextual factors, like location, physical activity, time of day, type of day, phone interaction, and phone battery level, are associated with receptivity to interventions. We also explored the relationship between receptivity and participant-specific characteristics, like age, gender, personality and device type, and receptivity. We further built machine-learning models to detect receptivity and achieved a 77\% improvement in F1-score over a biased random model.

{\color{red}Morrison~et~al.\  focused on a mobile stress-management intervention~\cite{morrison2017effect}. Their system randomly assigned study participants to one of three groups. Each group was using a different method for receiving push notifications. One group received the notifications occasionally (not daily), one group received them daily, and one group used a machine-learning model to receive `intelligent' notifications. However, the authors did not find any difference in how the participants interacted with notifications delivered at opportune times (as determined by their model) vs. those delivered at random times. We argue the negative result was because of the following challenges, and also highlight how we address those challenges in our work.

\begin{itemize}
    \item Based on prior work on interruptibility to generic phone notification, Morrison~et~al. used location labels (home, work, and other), movement from accelerometer, and time variables as the features~\cite{morrison2017effect}. While these features might be useful to detect interruptible moments for generic notifications, they might not translate to detecting receptivity to interventions.  In fact, our prior work showed that some features like location labels had no significant associations with receptivity to interventions. In comparison, we used more effective features that had shown prior associations with receptivity to interventions~\cite{kunzler:receptivity2019}.
    
    \item Morrison~et~al. built static personalized Naive Bayes models for each participant based on the data collected during an initial ``learning period.'' Since the model was static and did not update after the ``learning period,'' it is possible that their models did not have enough data during to be adequately trained. On the contrary, we used data from a prior study with 148 participants to build our static model, which resulted in a more robust and better-trained model.
\end{itemize}

Hence, we argue that our methodology of feature selection and model development could provide better insights towards the effectiveness of machine-learning models for detecting receptivity.  

}

In our current work, we build upon the prior research by deploying machine-learning models to detect receptivity in real-life situations. We used the data collected from our previous study~\cite{kunzler:receptivity2019} to build and deploy two machine-learning models to detect receptivity to physical-activity interventions. We deployed a static model that stayed unchanged through out the study, and an adaptive model that re-trained itself over the course of the study as participants interacted with the notifications. Our current work moves beyond ``post-study'' model-evaluations, and evaluates the effectiveness of deploying receptivity-detection models trained using data from a previous study.
    Further, this is the first work to \emph{deploy\/} an adaptive model to detect receptivity to JITAI and observe how the model performance changes as the study progresses. 

\section{Background} %
\label{sec:background}
In this section, we discuss the app used and dataset collected from our prior work by K\"{u}nzler~et~al.~\cite{kunzler:receptivity2019}, followed by the operationalization of \emph{receptivity}, and finally discuss how receptivity fits within the \ac{JITAI} framework.

\subsection{The Ally Study}
In an effort to understand receptivity to \ac{JITAI}s, in our prior work we developed a mobile \ac{JITAI} system to promote physical activity. The ``Ally'' app -- based on the open-source MobileCoach framework -- was a chat-based digital coach (for Android and iOS phones) that delivered an actual behavior-change intervention aimed at increasing the participant’s daily step count~\cite{kowatsch2017design, filler2015mobilecoach}. The previous study was conducted with 189 participants in Switzerland, over a period of 6 weeks. Participants received notifications that encouraged them to engage in conversation with the digital coach, which was a German-speaking chatbot motivating participants to increase their physical activity as measured by daily step count~\cite{kunzler:receptivity2019}.

Through that study, we reported interesting findings about the association between receptivity and participant-specific traits (like age, gender, device type and personality) and contextual factors (like battery level, device interaction, physical activity, location, and time of day). We also built several machine-learning models to infer different aspects of receptivity, and reported a 77\% increase in F1-score for detecting \emph{just-in-time\/} receptivity, over a biased random classifier (with a combination of participant specific traits and contextual factors), and a 50\% improvement in F1-score with contextual factors alone~\cite{kunzler:receptivity2019}.

Given the promising results in the post-study analyses, we decided to build upon that work by deploying in-the-moment receptivity-detection models to evaluate how these models perform in real-world situations.

Another reason to build upon that prior Ally study was the assumption we made about \emph{receptivity}. In the intervention design, all conversations started with a generic greeting message like ``Hello [participantName]'' or ``Good morning [participantName]''. Only when the participant responded to the greeting message, did the coach start sending the actual intervention messages. Hence, when we refer to receptivity to a message, we refer to receptivity to the \emph{initiating\/} message or ``start-of-conversation'' message.
Given that the Ally participants responded to initiating messages without looking at the actual intervention, we believe it might be possible to build models using the data collected in the Ally study and deploy it in a different study following a similar ``start-of-conversation'' message strategy.

\subsection{Receptivity within JITAI}
As proposed by Nahum-Shani~et~al., \ac{JITAI} has six key elements: a distal outcome, proximal outcomes, decision points, intervention options, tailoring variables, and decision rules~\cite{nahum2016just}. \emph{Distal outcome\/} is the ultimate goal the intervention is intended to achieve, \emph{proximal outcomes\/} are short-term goals the intervention aims to achieve, \emph{decision points\/} are points in time at which an intervention decision must be made, \emph{tailoring variables\/} are information concerning the individual that is used to decide when (i.e., under what conditions) to provide an intervention and which intervention to provide, \emph{intervention options\/} are an array of possible treatments/actions that might be employed at any given decision point, and \emph{decision rules\/} link the tailoring variables and intervention options in a systematic way.

Within this model, \emph{receptivity\/} can be considered as a \emph{tailoring variable}, one that indicates whether the user is available to receive an intervention at any given time. At a \emph{decision point}, the \emph{decision rules\/} can check the receptivity tailoring variable to choose from the various \emph{intervention options}, which could include postponing the intervention or delivering no interventions at that decision point. Hence, receptivity as a tailoring variable can help determine if, what, and when an intervention should be delivered.

\subsection{Operationalizing Receptivity}
Before discussing our methodology, it is important to establish precise metrics about what the models are trying to achieve. These definitions are consistent with the metrics we used in our prior work~\cite{kunzler:receptivity2019}.

\begin{itemize}
    \item \emph{Just-in-time response}: If a user views and responds to the initiating message within 10 minutes\footnote{We chose a 10-minute window to remain consistent with our prior work, where we also used a 10-minute window to define the receptivity metrics~\cite{kunzler:receptivity2019}. Further, prior work by Mehrotra~et~al. found that smartphone users accept (i.e., view the content of) over 60\% of phone notifications within 10 minutes of delivery, after which the notifications are left unhandled for a long time. They concluded that the maximum time a user should take to handle a notification arriving in an interruptible moment is 10 minutes~\cite{mehrotra2015designing}. Since ours is one of the early works to deploy receptivity detection models, we decided to follow this evidence and also use a 10-minute window.} of receiving the prompt, then the user is said to be in a receptive state and it counts as a `just-in-time response'.
    
    \item \emph{Response}: If the user responds to the initiating message at any time, even after the first 10 minutes, it counts as a `response'.
    
    \item \emph{Response delay}: the time (in seconds) elapsed between receipt of the initiating message and the user's first reply to it.

    \item \emph{Conversation engagement}: If the user replies to more than one message in a 10-minute window following the initiating message, it counts as `conversation engagement'.
    
\end{itemize}

In some contexts we aggregate these metrics over a period of time, e.g., over one day or over the duration of the study.
For a given period of time, the \emph{just-in-time response rate\/} is the fraction of initiating messages for which there was a \emph{just-in-time response}, the \emph{overall response rate\/} is the fraction of initiating messages for which there was a \emph{response}, the \emph{conversation rate\/} is the fraction of initiating messages that counted as \emph{conversation engagement}, and the \emph{average response delay\/} is the mean \emph{response delay}~\cite{kunzler:receptivity2019}.

\section{Approach} %
\label{sec:approach}

In this section, we discuss how we extended the Ally app, to fit our research goals, the models deployed in the app, and the study methodology.

\subsection{The \appname~app}

We modified the iOS version of the Ally app to create a new app we call \emph{\appname}. Similar to Ally, \appname~is a chat-based digital coach aimed at increasing daily step count. We show a screenshot of the app in Figure~\ref{fig:ally-screenshot}.
The app calculated a personalized step goal each day, based on the participants' $60^{th}$ percentile step count in the previous 9 days. The intervention components were chat-based conversational messages that were delivered by the digital coach and the participants had to choose from a set of pre-defined responses. The coach initiated the starting message of each conversation to each user at random times within certain time periods. In our study, we used three of the four intervention components used in the original Ally study~\cite{kunzler:receptivity2019,kramer2018investigating}:
(1)~The goal-setting prompt set the step goals for the day; it was delivered between 8~and 10~a.m daily.
(2)~The self-monitoring prompt informed participants about their progress and aimed to motivate them to complete their step goals; it was delivered at a random time between 10 a.m. and 6 p.m. to a randomly selected 50\% of participants each day.
(3)~The goal-achievement prompt informed the participants whether they achieved their goal for the day and aimed to motivate them to complete future goals; it was delivered at 9 p.m. daily.

\begin{figure}[h]
  \centering
  \includegraphics[width=0.5\linewidth]{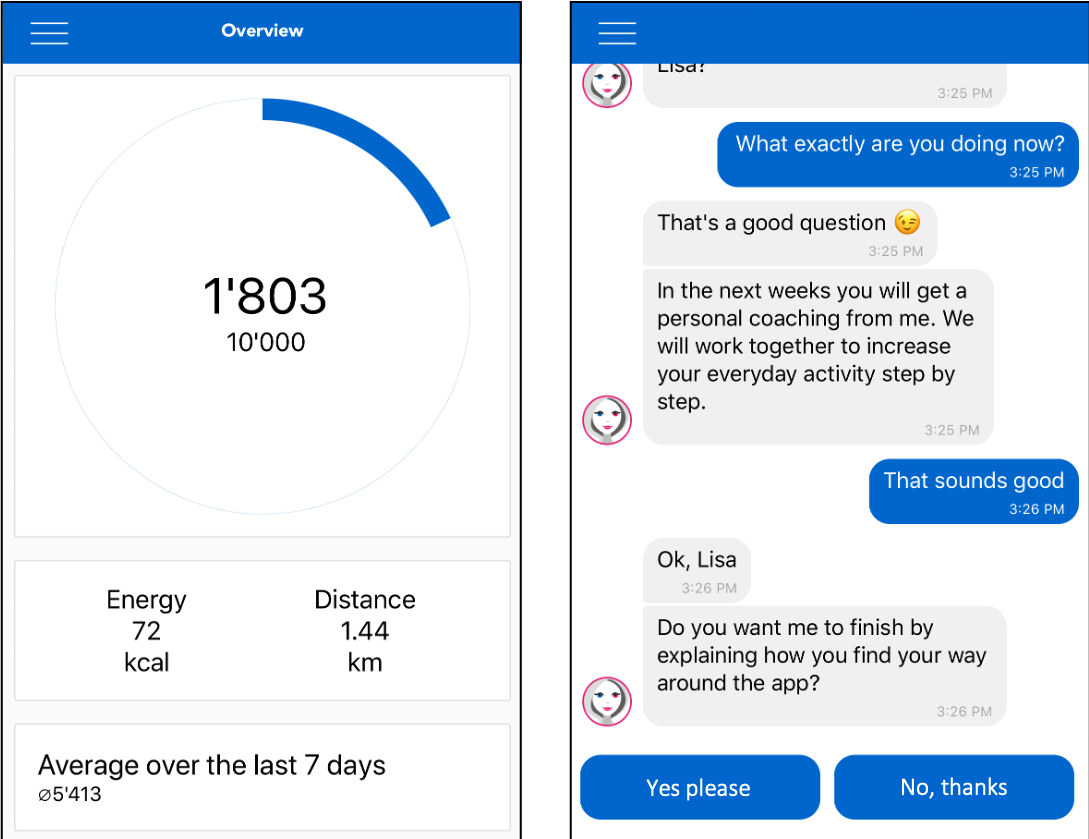}
  \caption{Two screenshots showing \appname's dashboard (left) and the chat screen for the interventions (right).}\label{fig:ally-screenshot}
\end{figure}

Further, \appname\  had a context-based receptivity module that continuously tracked several contextual features; \appname\  used this module to time the delivery of notifications, as follows.
For each day, for each participant, the server randomly chose three times (one in each of the three time blocks) to send a \emph{silent\/} push notification to that participant's app. 
When \appname~received the silent push from the server, it triggered the receptivity module to determine when to deliver that notification to the participant. 
During the first seven days, the receptivity module randomly selected either the control or static model, with equal weight. 
On the eighth day and after, the receptivity module randomly selected one of three models, with equal weight.
(The seven-day `warm-up' period allowed accumulation of participant-specific receptivity data before enabling the adaptive model.)
For each initiating message received, the app recorded which model was used to time its delivery~-- control, static or adaptive.
We detail the three models in Section~\ref{sec:building-models}.

\appname\  then delivered the notification about the initiation prompt if and only if the selected model inferred the user would be receptive at the current time.
The control module always agreed.
The static and adaptive models used their classifier to determine whether the current moment is `receptive'.
If the models did not find the current moment to be \emph{receptive},\footnote{A receptive moment is one at which the model predicts that the participant will respond to the initiating message within the next 10 minutes.} the app would try again by asking the same model every 5~minutes. If after 30 minutes the model never inferred an opportune moment, \appname\  delivered the notification on the $31^{st}$ minute;\footnote{{\color{red}It is important to note that the app only delivered \emph{one} prompt for each initiating message, either immediately after receiving the silent push from the server, or after 5, 10, 15, 20, 25, 30, or 31 minutes after the initial silent push, depending on the delivery mode and detection of receptive moments. There were no reminder prompts.}} in this case, it recorded the delivery mechanism as ``control'', since the notification was delivered at a random time, and not at an opportune moment.\footnote{It can be argued that a notification delivered at the $31^{st}$ minute might not be exactly like `control', since a machine-learning model has already passed it as non-receptive. We argue, however, that since the `control' model delivers notifications at random times, and since 31 minutes later than a random time is still a random time, it can be assumed that the `control' model delivered the notification. The goal of this work is to distinguish between moments identified as receptive by the machine-learning models and random moments. We argue all moments are either receptive or random, and the machine-learning models simply identify receptive moments from a series of random moments.}

We used the \appname~app to conduct a within subjects study with three experimental conditions for delivering the interventions: \textit{control}, \textit{static}, and \textit{adaptive}. It is important to note that the intervention delivery conditions did not affect the actual content of the interventions delivered by the app. 

Regardless of the chosen delivery model, the participant's response to any initiating message provided new data for use by the adaptive model.
There were three cases:
(a)~just-in-time response: the contextual state at the time of notification delivery was added with label `receptive';
(b)~later response: the contextual state at the time of notification delivery was added with label `non-receptive', and the contextual state at the moment of response was added with label `receptive' (since the participant was in a state-of-receptivity when they responded);
(c)~no response: the contextual state at the time of notification delivery was added with label `non-receptive'.
Whenever the adaptive model was selected as the delivery model, it first re-trained its model using any new data points added. 

We diagram the system design in Figure~\ref{fig:system-design}.
\begin{figure}[h]
  \centering
  \includegraphics[width=0.6\linewidth]{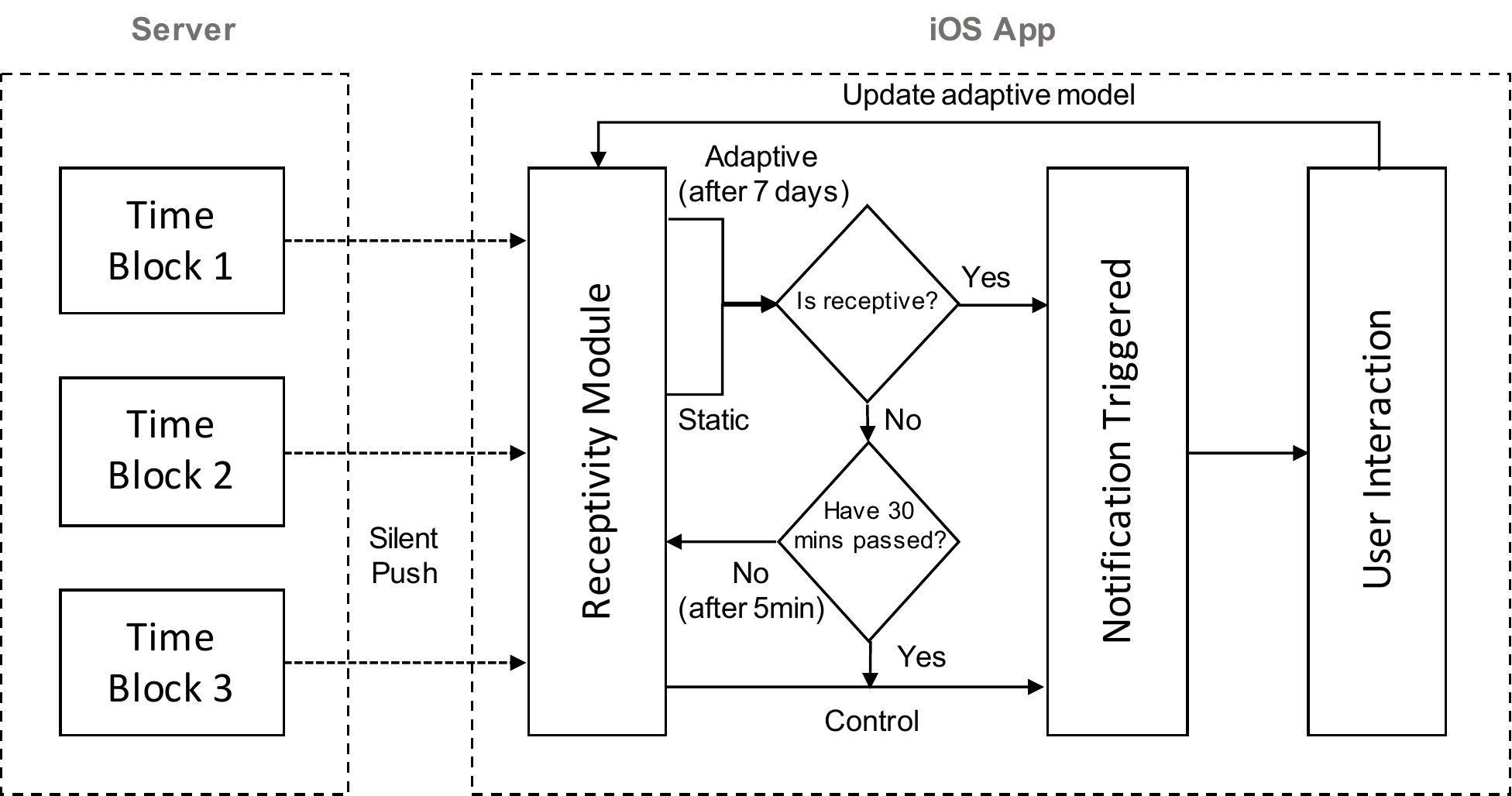}
  \caption{System design of the \appname~app.}\label{fig:system-design}
\end{figure}

\subsection{Building the Detection Models}
\label{sec:building-models}

We leveraged the data collected from iOS users in the previous Ally study to build new models that we deployed in \appname. We first discuss the features we selected to use in our models, followed by a description of the models themselves.

\subsubsection{Choice of Features}
In our prior work, we considered a variety of contextual features, like physical activity, device interaction, location type, type of day, time of day, and phone battery status~\cite{kunzler:receptivity2019}. Although most of those features can be calculated in real-time on an iOS device, one particular feature, \emph{location type\/} (like home, work or transit), is more complicated. To accurately compute the location type, algorithms need several weeks of location data to meaningfully cluster and derive categories of location (such as `home' or `work'). In our 2-3 week study there was insufficient data to derive the location-type feature; furthermore, we had found that (for Ally) iOS users' location type showed no significant associations with any receptivity metric~\cite{kunzler:receptivity2019}. We thus decided not to include location type as a feature in our models.

A list of features used in our models is shown in Table~\ref{tbl:features}.

\begin{table}[h]
\centering
\caption{List of features used in our models.}
    \resizebox{0.5\textwidth}{!}{%

\begin{tabular}{lll}
\toprule
\textit{Category}                            & \textit{Features} & \textit{Type}                                                                                         \\ \toprule
\multirow{2}{*}{\textbf{Date/Time}}          & Type of day       & \begin{tabular}[c]{@{}l@{}}Categorical \\ (weekday/weekend)\end{tabular}                              \\
                                             & Time of day       & \begin{tabular}[c]{@{}l@{}}Categorical \\ (morning, afternoon, evening)\end{tabular}                  \\ \hline
\multirow{2}{*}{\textbf{Phone Battery}}      & Battery Status    & \begin{tabular}[c]{@{}l@{}}Categorical\\ (charging, discharging, full)\end{tabular}                   \\
                                             & Battery Level     & \begin{tabular}[c]{@{}l@{}}Numerical\\ (1\%-100\%)\end{tabular}                                                                                              \\ \hline
\multirow{3}{*}{\textbf{Device Interaction}} & Lock State        & \begin{tabular}[c]{@{}l@{}}Categorical\\ (locked, unlocked)\end{tabular}                             \\
                                             & Lock change time  & \begin{tabular}[c]{@{}l@{}}Numerical\\ (in seconds)\end{tabular}                                                                                              \\
                                             & Wi-Fi connection  & \begin{tabular}[c]{@{}l@{}}Categorical\\ (connected, disconnected)\end{tabular}                       \\ \hline
\textbf{Activity}                            & Physical Activity & \begin{tabular}[c]{@{}l@{}}Categorical\\ (still, on foot, on bike, running\\ in vehicle)\end{tabular} \\ \bottomrule
\end{tabular}
}
\label{tbl:features}
\end{table}

\subsubsection{The \emph{Static\/} and \emph{Adaptive\/} Models}

We implemented two machine-learning models in \appname, our iOS app.
We trained the \textit{static model\/} before deployment (using data from the previous Ally study) and used it, unchanged, for all participants and all days throughout the study.
The \textit{adaptive model\/} used the receptivity data of individual participants as they progressed through the study; 
it was re-built (within the app) every time a new receptivity the system trigerred the adaptive model. 

Both these models were trained to predict \emph{just-in-time response}.
While we use several metrics of receptivity in our work, the main emphasis is on the presence of a just-in-time response. For completeness, however, we report the effect of our models on the various receptivity metrics.

\vspace{0.8ex}\noindent
We next provide the details of each model.
\vspace{1.0ex}

\emph{Static Model}: We used CoreML to build and integrate the static model with the iOS app~\cite{apple:coreml}. 
We split the original Ally iOS data (with 141 users) into five equal non-overlapping groups. We used \ac{LOGO} cross-validation to evaluate two built-in models within CoreML~-- MLRandomForestClassifier and MLSupportVectorClassifier. These classifiers are CoreML's implementation of RandomForest and SVM, respectively.

We tuned the models to have higher recall, since we wanted \appname\ to recognize most opportune moments, even if it was at the cost of precision. We compared the models with a random classifier as a baseline and chose the model that demonstrated a greater improvement in F1 score. The SVM classifier achieved a mean F1 score of 0.36, whereas the random baseline classifier achieved only F1 score of 0.25, which is an improvement of 40\% over the baseline. The RandomForest classifier achieved a mean F1 score of 0.33, only 32\% improvement over baseline. We thus chose the SVM classifier as the static model to be included in our app. We show the comparison between Precision, Recall and F1 score of these three models in Figure~\ref{fig:training-comparison}.

\begin{figure}[htb]
  \centering
  \includegraphics[width=0.45\linewidth]{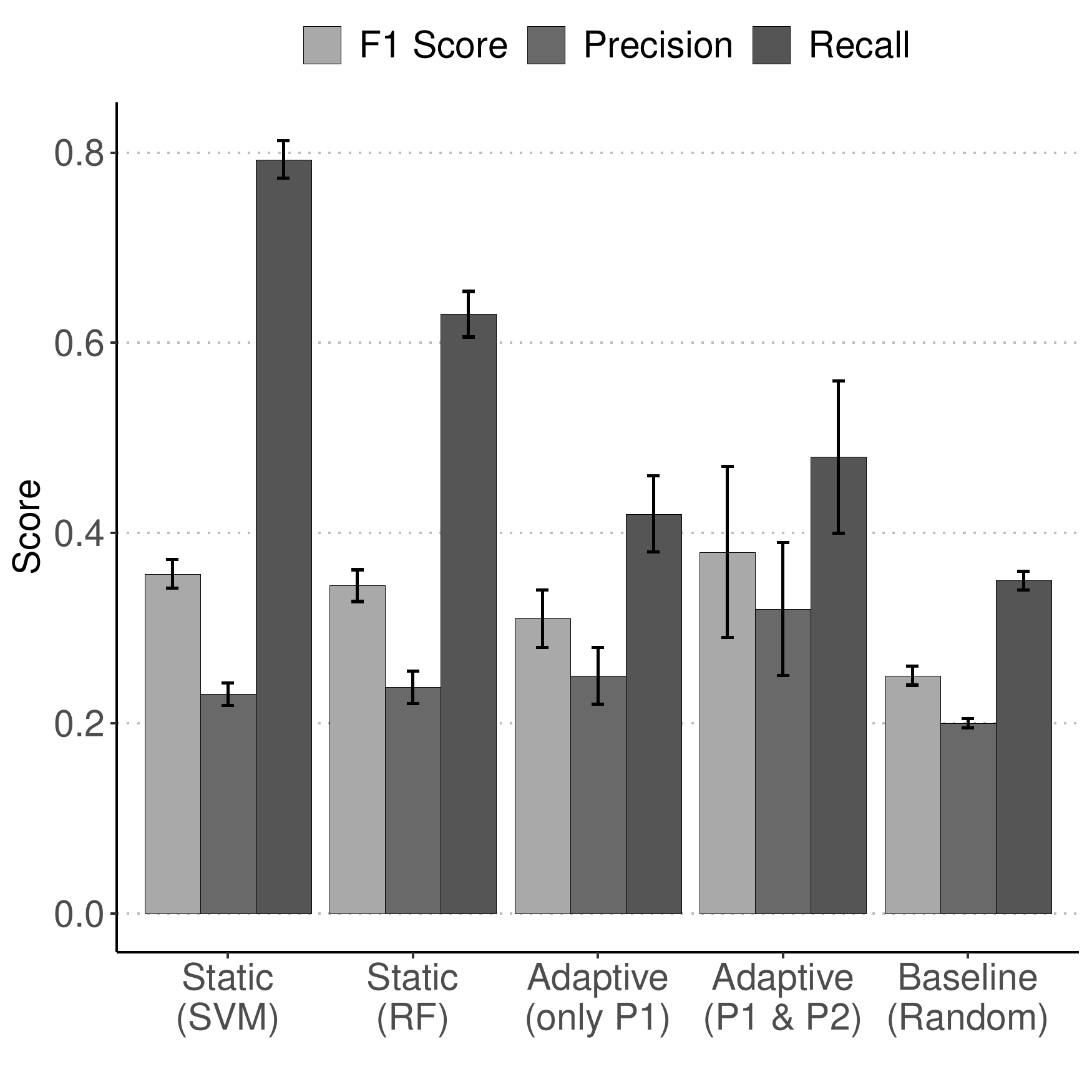}
  \caption{\color{red}Comparison of the different models. All models were evaluated using 5-fold \ac{LOGO} cross-validation }\label{fig:training-comparison}
\end{figure}

\emph{Adaptive Model}: Implementing the adaptive model was more complicated because we needed to re-build the model every time a new receptivity data point was available. In iOS 13, Apple added functionality to update CoreML models on-the-fly. Because iOS 13 was released just a couple of months before our study, however, we chose not to depend on this new feature of CoreML.  To do so may have unnecessarily narrowed our potential participant population by excluding those who had not yet updated to the latest iOS version. 
We instead used an open-source library called AIToolBox~\cite{aitoolbox}. This library, however, had not been modified in a few years, so we had to make significant changes to make it compliant with the newer versions of iOS and Swift.
Since the adaptive model would be re-trained for every notification, we wanted a model that would not be resource-intensive while being re-trained. Based on preliminary tests, \ac{LR} had the fastest training time on the device, without significantly sacrificing detection performance.

In the adaptive model, the participant's recent receptivity data was added to the model's training dataset to help with future detection. Given the structure of our study, however, each participant was prompted at most three times per day and there were thus few data points even after seven days.  We thus followed a `dual-model' approach: the adaptive model's output probability was the average of the output probability from `P1', an \ac{LR} model trained on data from the prior Ally study, and `P2', an \ac{LR} model trained on the participant's personalized data accumulated thus far.\footnote{We explain the rationale for a dual-model approach in the supplementary document.} If the output probability was greater than $0.50$, the adaptive model classified that instance as `receptive'. This dual-model approach enabled us to introduce a degree of personalization without being concerned about high variance of the personalized model developed from a limited set of data points.

{\color{red}As ours is one of the first works to \textit{deploy} adaptive models to detect receptivity, we made several decisions based on what we thought might work. We considered the possibility of adding data to a single batch and training a single model. In our case, the base P1 model had 600 data points. Based on our study design, over the course of the study, each participant was expected to receive between 40-50 initiating messages. Since, at best, there would be fewer than 10\% participant-specific data points in our base model, we did not know the extent to which a single model would learn participant-specific characteristics from the data. Hence, we decided to use a dual-model approach. 
   
   To assign weights to the individual models, a sophisticated approach would be to initially start out with a low weight on the personalized component and increase the weights as data is accumulated over time. This method, however, needed more decisions to be made, like what the starting weights should be, or whether weights should vary based on the number of days in the study or the number of available instances, or what degree should the weight vary with new data. Lack of any prior work in this domain made these decisions non-trivial. Hence, as an exploratory evaluation we decided to start with fixed weights for each model for the duration of our study. Further, due to lack of a better choice, we assigned a weight of 0.5 to each model. Our rationale was that equal weights would help handle disagreements between the models, e.g., if P2's output probability was 0.90, and the P1's output probability was 0.20 (or vice-versa), then equal weights ensured that adaptive model would trigger an alert.}

We trained the P1 model using data from the previous Ally study. To ensure the model was light enough to run on the phone, we had to under-sample our training data.\footnote{\color{red}Unlike CoreML, which we used for the static model, the AIToolBox library does not export a pre-built model. The library trains and re-builds the model every time classification tasks need to be performed. Hence we had to include a trimmed version of the training data in the app so that the P1 model could train itself.} We chose the Instance Hardening Threshold (IHT) method, which generates a balanced under-sampled dataset by eliminating instances that are frequently misclassified, i.e., have high instance hardness~\cite{smith:instance-hardness14}. We evaluated the P1 model by LOGO cross-validation with the same non-overlapping groups we used for evaluating the static model. See Figure~\ref{fig:training-comparison}; P1 achieved an F1 score of 0.31, which is slightly lower than the F1 score for the static model, perhaps because P1 used \ac{LR} whereas the static model used SVM, or perhaps because we had to train P1 on a trimmed dataset. 

{\color{red}To evaluate the effectiveness of P2, we ran simulations over the previously collected Ally data. During each fold of the 5-Fold cross-validation, we used the 20\% users data to simulate the effectiveness of the P2 model. We used the P1 model trained on the 80\% of users in each fold. We evaluated P2 on the entire study duration of each participant, which was a maximum of 6 weeks. It is worth noting that while we conducted the previous Ally study for 6 weeks, there were several participants who dropped out before completing the the full 6 weeks.
The new result is shown as `Adaptive (P1 \& P2) in Figure~\ref{fig:training-comparison}. We observe that the complete adaptive model led to an F1 score of 0.38, which is a 52\% improvement over the random baseline. However, there is much higher deviation in the Adaptive model as compared to the other models. We believe this is because of the difference in the participation duration of each participant, thus resulting in varied performance of the P2 model. }

Please note the purpose of deploying both static and adaptive models was not to compare these models with each other, but instead to observe (a)~how did each model perform compared to the control model, and (b)~how did the performance of the adaptive model change over time?

\subsection{Study Logistics and Procedure}
Unlike the Ally app, the \appname~app was released only for iOS users. Also, instead of releasing it through the App Store, we used Apple's in-house distribution program to distribute the apps to the participants, who could download the app by navigating to a specific webpage. Since the goal of the study was to evaluate participant receptivity, we did not want to bias the participant's interaction and usage of the app by providing monetary incentives for using the app or for engaging with the app. Instead, our study strategy used `deception' to mask the actual goals of the study. During recruitment, we told the participants the goal of the study was to understand how different contexts affect the physical activity levels of a person throughout their day. We asked participants to interact naturally with the \appname~app and compensated them the equivalent of USD~25 if they installed the app for at least two-thirds of the study duration, i.e., 14 days.

The study protocol (including the use of deception) was approved by the \ac{IRB} of the respective institutions. As required by the \ac{IRB}, at the end of the 3-week period we emailed the participants informing them of the real goal of the study with an explanation of why deception was needed.

We used Facebook advertisements to reach potential participants, with the hope of reaching a diverse participant pool. Our search criteria was set to adults over 18 years, and belonging to a single timezone (due to technical limitations). 
Participants who clicked on the advertisement were taken to a landing page where we explained the study; and if interested, people could digitally sign the consent form. Once the consent form was signed we emailed the app download link and instructions to the prospective participants. The Facebook advertisement had a reach of over 30,000 people, out of which over 750 participants were redirected to our landing page; of those, 189 interested people filled out the consent form, of which 83 users downloaded the app and started the intervention. Of the 83 participants, 64 were female and 19 were male. The median age was 30 years $\pm 10.8$ years. We had a staggered recruitment approach, and not all participants had the same start and end dates. While each participant was enrolled in the study for 3~weeks, our total data-collection period spanned a period of 6~weeks (between September 2019 and November 2019).
We show a detailed demographic breakdown of participants along with their average response rates in Table~\ref{tbl:demographics}.

\begin{table}[htb]
  \caption{Participant demographics with average response rates.}\label{tbl:demographics}
      \resizebox{0.6\textwidth}{!}{%

  \begin{tabular}{lcccc}
    \toprule
    &Male&Female&Undisclosed&Total\\
    \midrule
    Showed interest in participating & $47$ & $142$ & $1$ & $190$\\
    \hline
    Installed the app & $19$ & $64$ & $0$ & $83$\\
    Age & $34 \pm 11.7$ & $30 \pm 10.6$ & & $30 \pm 10.8$\\
    Overall response rate & $69.45\%$ & $70.67\%$ & & $70.39\%$\\
  \bottomrule
\end{tabular}
}
\end{table}
Prior research papers on `receptivity for interventions' and `interruptibility to phone notifications' have defined a `participant selection criteria' and included only data from those selected participants in their analysis~\cite{kunzler:receptivity2019, choi:receptivity19, pielot2017beyond, mehrotra2016my}, like response to at least 14 questionnaires~\cite{mehrotra2016my}, or data collection for at least 10 days and 20 responses~\cite{pielot2017beyond}, or top 95 percentile of users who responded~\cite{kunzler:receptivity2019}. In our work, however, we avoid use of any such selection criteria and report results for data from all the 83 users for most of the analysis. Across the 83 users, we had 1091 messages delivered by the \emph{control\/} model, 691 messages delivered by the \emph{static\/} model, and 241 messages delivered by the \emph{adaptive\/} model; resulting in a total of 2023 delivered messages. We present a distribution of the number of times each model was used to deliver a message over the course of the study in Figure~\ref{fig:condition-distribution}.

\begin{figure}[htb]
    \centering
  \includegraphics[width=0.7\linewidth]{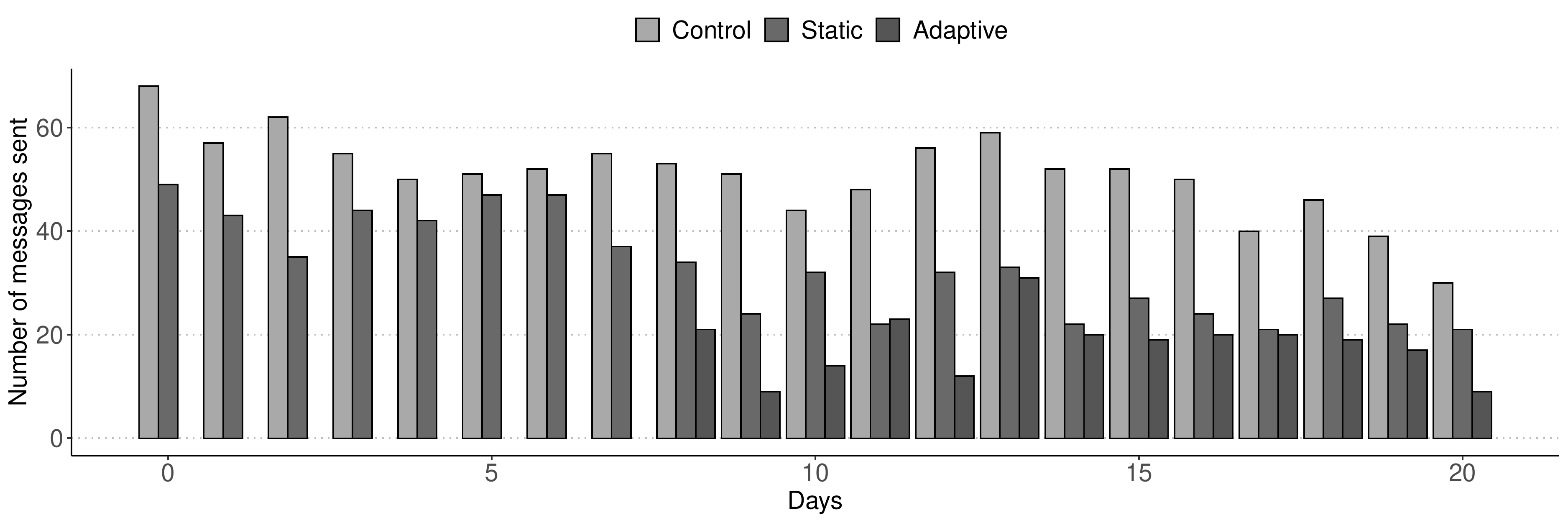}
  \caption{Daily distribution of the models used to trigger intervention alerts.}\label{fig:condition-distribution}
\end{figure}

\subsection{Problem with Deployment}
\label{sec:field-problem}
Approximately two or three weeks from the start of data collection we realized the adaptive model was \emph{never\/} getting triggered for participants. Even after the initial 7~days, the \appname~app was triggering only the control and static models, because a bug was preventing the app from triggering the adaptive model. We fixed the bug, and released an update and asked all users to update to the new version. Unfortunately some users had already completed their study duration, whereas others did not update the app. In the end, we eventually had only 61 users who received one or more initiating messages through the adaptive model. For these 61 users, we had 785 messages delivered by the control model, 541 messages delivered by the static model, and 241 messages delivered by the adaptive model, leading to a total of 1567 delivered messages.\footnote{\color{red}Given our intervention design, 1567 messages might seem less for 61 participants receiving and average of 2.5 messages per day for 21 days. However, we had several participants who dropped out over the course of the study. Since we did not want to enforce an arbitrary ``inclusion criteria'' -- based on the number of days or the number of messages -- we decided to follow an intention-to-treat approach and included all participants who installed the app and started the study in our analysis.}

\section{Our Hypotheses}
Before evaluating the results, we formed the following hypotheses based on our expectation of the outcomes if the models performed as intended.
\vspace{0.8ex}

\noindent
For \emph{RQ-1},
\begin{itemize}
    \item \emph{H-1}: We hypothesize that on a population level, across all initiating messages, there would be a significant difference in receptivity across the three delivery models. The \emph{static\/} and \emph{adaptive\/} models should both have significantly higher receptivity than the \emph{control\/} model.

\end{itemize}

\noindent
For \emph{RQ-2},
\begin{itemize}
    \item \emph{H-2}: We hypothesize that individual participants would have higher receptivity when interventions were delivered through the \emph{static\/} and \emph{adaptive\/} models, as compared to the \emph{control\/} model.
\end{itemize}

\noindent
For \emph{RQ-3},
\begin{itemize}
    \item \emph{H-3}: We hypothesize that the receptivity to interventions delivered by the \emph{static\/} model would remain constant over the course of the study.
    
    \item \emph{H-4}: We hypothesize that the receptivity to interventions delivered by the \emph{adaptive\/} model would increase as the study progressed.
\end{itemize}

Note that while the hypotheses mention \emph{receptivity}, our primary focus is on \emph{just-in-time response}. For completeness, however, we also report the other metrics of receptivity, which we term ``secondary metrics''. Hence, based on the type of metric, ``better'' receptivity would mean an increase in just-in-time response rate, increase in response rate, increase in conversation engagement rate, and a decrease in response delay.

\section{Evaluation} %
\label{sec:evaluation}
In this section we analyze the receptivity data and evaluate our hypotheses. For clarity, we break the evaluation into three parts, to reflect the three research questions we seek to answer. Note, we adjusted all the $p$-values to account for multiple comparisons: we used the Benjamini-Hochberg (BH) procedure~\cite{benjamini1995controlling} to correct for the expected proportion of Type I errors (or False Discovery Rate) across all hypotheses.

\subsection{Exploring \emph{RQ-1}}
\begin{quote}
\textit{\RQone}
\end{quote}

We start by analyzing the receptivity metrics for each initiating message. We fit a binomial \ac{GLM} to evaluate the associations the \ac{ML} models had on just-in-time response as compared to the control model.\footnote{We choose the appropriate test based on the type of the dependent variable. We used a binomial \ac{GLM} for the just-in-time response, response, and conversation-engagement metrics. We used a one-way \ac{ANOVA} for response delay. From a typical statistical analysis perspective, it may initially seem that a population level analysis (across all intervention messages) violates the independence assumptions of GLM. However, it is important to consider the research question we aim to answer, i.e., how did the \textit{static} and \textit{adaptive} models perform when compared to the control model, across all intervention prompts. We argue that, from a machine-learning model's perspective, each new prediction is independent of prior predictions and hence can be considered independent. Thus, in this regard we are not violating the assumption for GLM. Such population level analyses is consistent with prior works in receptivity~\cite{kunzler:receptivity2019} and interruptibility~\cite{mehrotra2016my, mehrotra2017understanding}.} {\color{red}We found that the different delivery modes (control, static, and adaptive)} had a significant effect on the just-in-time response ($\chi^2$(2) = $23.189$, $p<0.001$).\footnote{The Chi Square test compares the GLM to a null model.} On post-hoc analysis with Dunnett’s Test, we observed that the static model showed a significant improvement of over 38\% in just-in-time receptivity when compared to the control model ($p<0.001$). The adaptive model had an improvement of more than 15\% over the control model, but the improvement was not significant ($p=0.271$). 

Next, we discuss the secondary metrics; we observed that the type of machine-learning model had a significant effect on the likelihood of ``response'', i.e., if the participant ever responded to the initiating message (irrespective of time), $\chi^2$(2) = $15.001$, $p=0.003$. Post-hoc analysis with Dunnett's Test revealed that, when compared with the control model, both the static and adaptive models led to a significant increase in likelihood of response, of approximately 12\% each ($p=0.003$ and $p=0.048$, respectively). This is an interesting observation as it suggests that participants were more likely to respond to initiating messages that were generated through the static and adaptive models as compared to the control model, even though the adaptive model did not lead to a significantly higher likelihood of just-in-time response. Further, the type of machine-learning model showed a significant effect on conversation engagement, $\chi^2$(2) = $18.741$, $p=0.001$. Post-hoc analysis revealed that the static model led to a 34\% increase in the likelihood of conversation engagement relative to the random model ($p<0.001$). As in the case of just-in-time response, the adaptive model did not result in a significant improvement in conversation engagement, with an increase of 12\% over the control model ($p=0.439$). Finally, for response delay, one-way \ac{ANOVA} did not reveal a significant effect of model type, $F(2,1382) = 1.576, p = 0.310$; we nonetheless conducted Dunnett's test to observe the differences. Although the difference was not statistically significant, the static model -- on average -- led to a slightly shorter response delay of 16 minutes, about 17\% faster than in the control model. The adaptive model did not show any differences as compared to the control model.
We present the detailed results in Table~\ref{tbl:message-level}.

\begin{table}[htb]
\centering
\caption{Detailed analysis across all initiating messages. We report the absolute change of the static and dynamic models over the control model, along with the percentage improvement in brackets.}
\label{tbl:message-level}
    \resizebox{0.9\textwidth}{!}{%

\begin{tabular}{lrrcrcl}
\toprule
\multirow{2}{*}{Comparison} & \multirow{2}{*}{\begin{tabular}[c]{@{}c@{}}Mean Difference\\ (\% change)\end{tabular}} & \multirow{2}{*}{Std. Error} & \multicolumn{2}{c}{95\% Confidence Interval}                & \multicolumn{2}{c}{\multirow{2}{*}{Adj. $p$-value}} \\ \cline{4-5}
                            &                                  &                             & Lower Bound                & \multicolumn{1}{c}{UpperBound} &                                      &     \\ \hline
\hline
\multicolumn{7}{l}{\textbf{Just-in-time response} (as likelihood; control $=0.284$)}                                                                                                                                           \\ \hline \hline
\emph{static} -- \emph{control}           & +0.108 (+38.02\%)                            & 0.022                       & \multicolumn{1}{r}{0.057}  & 0.164                          & \multicolumn{1}{r}{\textless{}0.001} & *** \\
\emph{adaptive} -- \emph{control}         & +0.044 (+15.49\%)                           & 0.033                       & \multicolumn{1}{r}{--0.034} & 0.122                          & \multicolumn{1}{r}{0.270}            &     \\ \hline
\hline
\multicolumn{7}{l}{\textbf{Overall response} (as likelihood; control $=0.656$)}                                                                                                                                           \\ \hline \hline
\emph{static} -- \emph{control}           & +0.080 (+12.19\%)                            & 0.022                       & \multicolumn{1}{r}{0.028}  & 0.133                          & \multicolumn{1}{r}{0.003} & ** \\
\emph{adaptive} -- \emph{control}         & +0.077 (+11.73\%)                            & 0.032                       & \multicolumn{1}{r}{0.006} & 0.154                          & \multicolumn{1}{r}{0.048}            &   *  \\ \hline
\hline
\multicolumn{7}{l}{\textbf{Conversation engagement} (as likelihood; control $=0.268$)}                                                                                                                                           \\ \hline \hline
\emph{static} -- \emph{control}           & +0.093 (+34.70\%)                            & 0.022                       & \multicolumn{1}{r}{0.045}  & 0.150                          & \multicolumn{1}{r}{\textless{}0.001} & *** \\
\emph{adaptive} -- \emph{control}         & +0.030 (+11.19\%)                            & 0.032                       & \multicolumn{1}{r}{--0.044} & 0.109                          & \multicolumn{1}{r}{0.439}            &     \\ \hline
\hline
\multicolumn{7}{l}{\textbf{Response delay} (in minutes; control $=90.530$)}                                                                                                                                           \\ \hline \hline
\emph{static} -- \emph{control}           & --16.670 (--18.41\%)                            & 9.465                       & \multicolumn{1}{r}{--38.870}  & 5.532                          & \multicolumn{1}{r}{0.144} &  \\
\emph{adaptive} -- \emph{control}         & --4.006 (--4.42\%)                            & 13.733                       & \multicolumn{1}{r}{--36.227} & 28.213                          & \multicolumn{1}{r}{0.819}            &     \\ \bottomrule
\multicolumn{7}{l}{\textbf{.} $p<0.1$, * $p<0.05$, ** $p<0.01$, *** $p<0.001$\hfill} \\ \bottomrule
\end{tabular}}
\end{table}

\subsection{Exploring \emph{RQ-2}}
\begin{quote}
\textit{\RQtwo}
\end{quote}

While investigating the hypotheses for \emph{RQ 1}, we found that overall, across the population, the static model led to a significant improvement in just-in-time response, overall response, and conversation engagement. While such an analysis provides a good representation of receptivity across all users, our second research question aimed to evaluate \emph{within-participant\/} differences when they received a static or adaptive model as compared to the control model. The goal is to evaluate whether an individual participant's receptivity changed based on the model used to deliver the initiating message.

We used generalized linear mixed effects models for our analysis. Since the goal was to observe how individual participant's receptivity changed based on the model used to deliver the message, each user should receive at least one message through each model for a proper comparison. As noted in Section~\ref{sec:field-problem}, however, a software problem meant only 61 participants received at least one initiating message through the adaptive model. Hence, for this analysis, we included data only from these 61 participants.

We observed that the model type had a significant effect on the just-in-time response rate ($\chi^2$(2) = $13.433$, $p=0.001$). On post-hoc analysis, we observed that the static model showed a significant improvement of over 36\% in just-in-time receptivity when compared to the control model ($p=0.002$). This result suggests that if a participant received a prompt from the static model, they were more likely to be receptive than if the same participant received the prompt through the control model. The adaptive model led to an increase of almost 10\% over the control model, but the result was not significant ($p=0.558$). 

For the secondary metrics, we observed that the type of model had an effect on the likelihood of response ($\chi^2$(2) = $8.364$, $p=0.00$). Post-hoc analysis revealed that only the static model had a significant improvement over the control model, with an improvement of almost 10\% ($p=0.015$). Further, our analysis showed that the model type had a significant effect on the likelihood of conversation engagement ($\chi^2$(2) = $10.407$, $p=0.017$), with post-hoc analysis revealing that the static model led to an improvement of over 32\% in the likelihood of conversation engagement over the control model ($p=0.007$). Finally, using a repeated measures \ac{ANOVA}, we did not find any significant effect of model type on the response delay. Although not statistically significant, the static and adaptive model led to a 20\% and 13\% reduction in time taken to respond to interventions, respectively. We present the detailed findings in Table~\ref{tbl:within-user}.

Further, when we added a random slope for `model type' to the mixed effects model, a comparison between the two did not reveal any difference ($p=1.00$), thus suggesting that the within-participant differences were similar across all participants. These results are promising, and indicate that individual participants were more receptive when they received prompts by the static model as compared to the control model.

{
\color{red}
\begin{table}[htbp]
\centering
\caption{Detailed analysis to understand \emph{within-participant\/} differences. We report the absolute change of the static and dynamic models over the control model, along with the percentage improvement in brackets.}
\label{tbl:within-user}
\begin{tabular}{lrrrrrl}
\toprule
\multirow{2}{*}{Comparison} &
  \multicolumn{1}{c}{\multirow{2}{*}{\begin{tabular}[c]{@{}c@{}}Mean Difference\\ (\% change)\end{tabular}}} &
  \multicolumn{1}{l}{\multirow{2}{*}{Std. Error}} &
  \multicolumn{2}{c}{95\% Confidence Interval} &
  \multicolumn{2}{c}{\multirow{2}{*}{Adj. $p$-value}} \\ \cline{4-5}
 &
  \multicolumn{1}{c}{} &
  \multicolumn{1}{l}{} &
  \multicolumn{1}{c}{Lower Bound} &
  \multicolumn{1}{c}{Upper Bound} &
  \multicolumn{2}{c}{} \\ \hline \hline
\multicolumn{7}{l}{\textbf{Just-in-time response} (as likelihood; control = 0.276)}        \\ \hline \hline
static -- control   & +0.101 (+36.60\%)  & 0.033  & 0.035   & 0.170  & 0.002 & **  \\
adaptive -- control & +0.027 (+9.58\%)   & 0.041  & --0.044  & 0.109  & 0.558 &  \\ \hline \hline
\multicolumn{7}{l}{\textbf{Overall response} (as likelihood; control = 0.738)}             \\ \hline \hline
static -- control   & +0.072 (+9.75\%)   & 0.028  & 0.015   & 0.116  & 0.015 & *  \\
adaptive -- control & +0.031 (+4.20\%)   & 0.038  & --0.046  & 0.092  & 0.493 &  \\ \hline \hline
\multicolumn{7}{l}{\textbf{Conversation engagement} (as likelihood; control = 0.261)}      \\ \hline \hline
static -- control   & +0.084 (+32.18\%)  & 0.034  & 0.021   & 0.153  & 0.007 & ** \\
adaptive -- control & +0.009 (+3.44\%)   & 0.040  & --0.057  & 0.089  & 0.819 &  \\ \hline \hline
\multicolumn{7}{l}{\textbf{Response delay} (as minutes; control = 99.500)}                 \\ \hline \hline
static -- control   & --19.950 (--20.05\%) & 11.725 & --39.500 & 3.500  & 0.124 &  \\
adaptive -- control & --13.830 (--13.89\%) & 13.585 & --41.000 & 13.500 & 0.439 &  \\ \bottomrule
\multicolumn{7}{l}{\textbf{.} $p < 0.1$, * $p<0.05$, ** $p<0.01$, *** $p<0.001$\hfill}      \\ \bottomrule
\end{tabular}
\end{table}
}

Given our inclusion criteria for this analysis, we have some users who received just one message from the adaptive model. It can be argued that including such users might introduce some outliers in our data, which might affect the analyses. However, given the lack of a better (and scientifically justifiable) inclusion criteria, we nonetheless included users who received only one message. To explore how a more stringent criterion would affect the results, we briefly analyzed the data from participants who received at least 5 messages in each category. Doing so ensured that we only included participants who were active (or stayed) in the study for a longer duration. We observed that the static model led to a significant increase in the just-in-time response rate, over 40\% higher than the control model. This result suggests that the difference in models becomes more prominent as we evaluate the more ``active'' users. In a way, our results (presented in Table~\ref{tbl:within-user}) can be thought of as the ``worst-case'' results.

\subsection{Exploring \emph{RQ-3}}
\begin{quote}
\textit{\RQthree}
\end{quote}

As we report in the preceding sections, the messages delivered by the static model led to receptivity metrics that were significantly higher than those the control model, for most situations. The adaptive model, however, did not seem to perform significantly better. Those results were based on an analysis across the full study period. A \emph{day-by-day analysis}, however, may provide more insights regarding whether and how the adaptive model's performance changed over the days~-- in short, whether it adapted well to each participant. This analysis also helps to evaluate our third research question. Given the nature of the adaptive model, we expected it to improve over time as more individual-specific data was added to the model.

{\color{red}To this end, we added a new variable -- the participants' day in study (from day~1 to day~21) -- as an interaction effect to the binomial generalized linear mixed effects model used for RQ-2. To best understand and vizualize the results, we plot the effects of the model types over time as estimated from mixed effect model in Figure~\ref{fig:overtime}. While visualizing the trends, it is important to note that the confidence interval for the adaptive model is quite wide for the first few days, because the adaptive model was not triggered till day 8 and hence no actual data points for the adaptive model during that period.

As the study progressed, we found that the just-in-time response rate dropped significantly for the control model ($p=0.011$) (Figure~\ref{fig:overtime}a). For the static model, there was a slight downward trend, but it was not significant. For the adaptive model there was a steep upward trend with a slope of $0.0092$, suggesting that the just-in-time response to adaptive model increased by almost 1 percentage-point each day; this trend was not statistically significant ($p=0.287$). The observation -- although not statistically significant -- is encouraging, suggesting that the adaptive model was able to learn and personalize over time, and eventually improving the just-in-time response. In fact, after day 19, the adaptive model seems to have had higher just-in-time response rate than the static model. Further, on Day 21, the adaptive model had an increase of over 51\% in just-in-time response rate as compared to Day 8.

\begin{figure}[]%
    \centering
    \subfloat[Effect of the different models on the just-in-time response rate over time.]{{\includegraphics[width=0.4\linewidth]{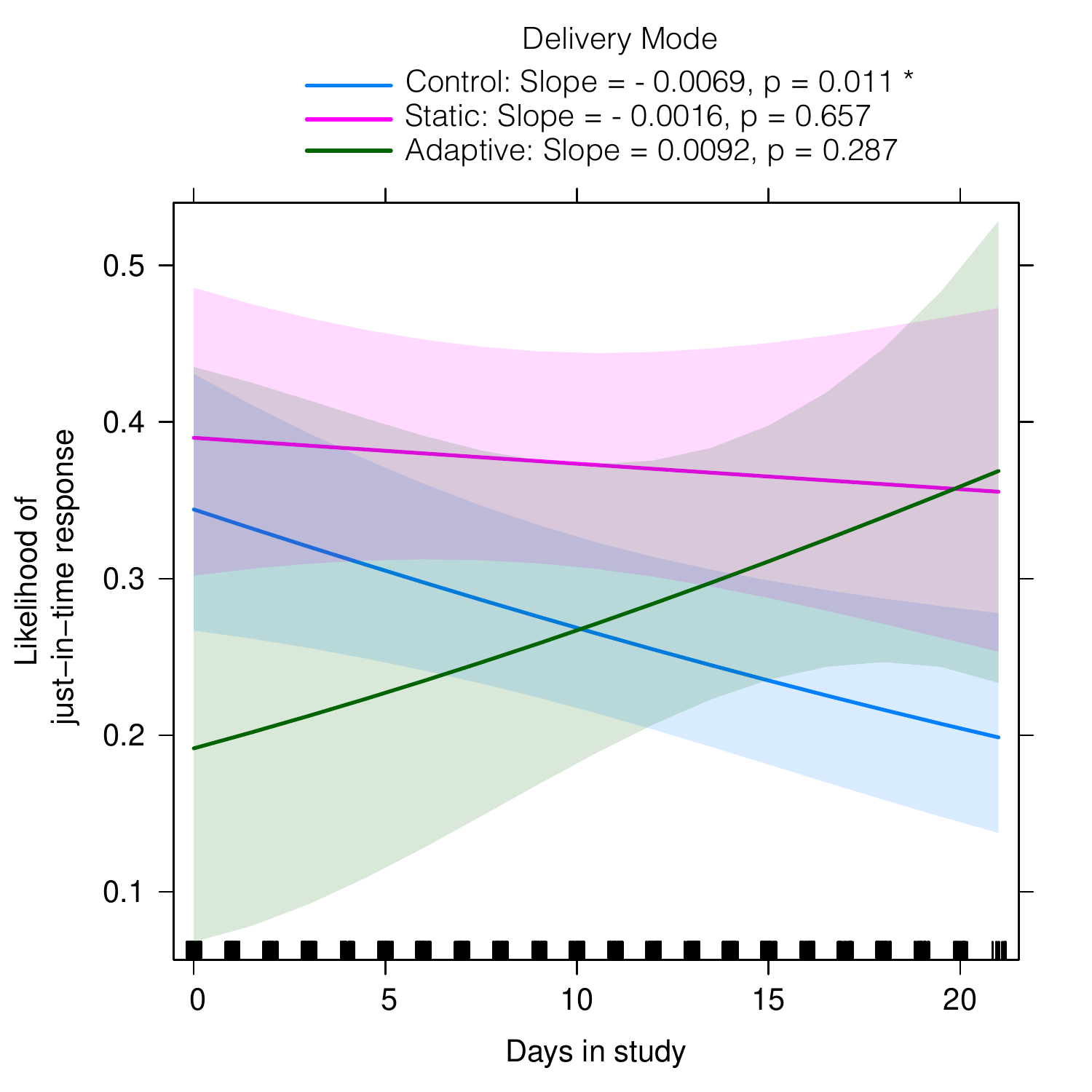} }}%
    \qquad
    \subfloat[Effect of the different models on the conversation-engagement rate over time.]{{\includegraphics[width=0.4\linewidth]{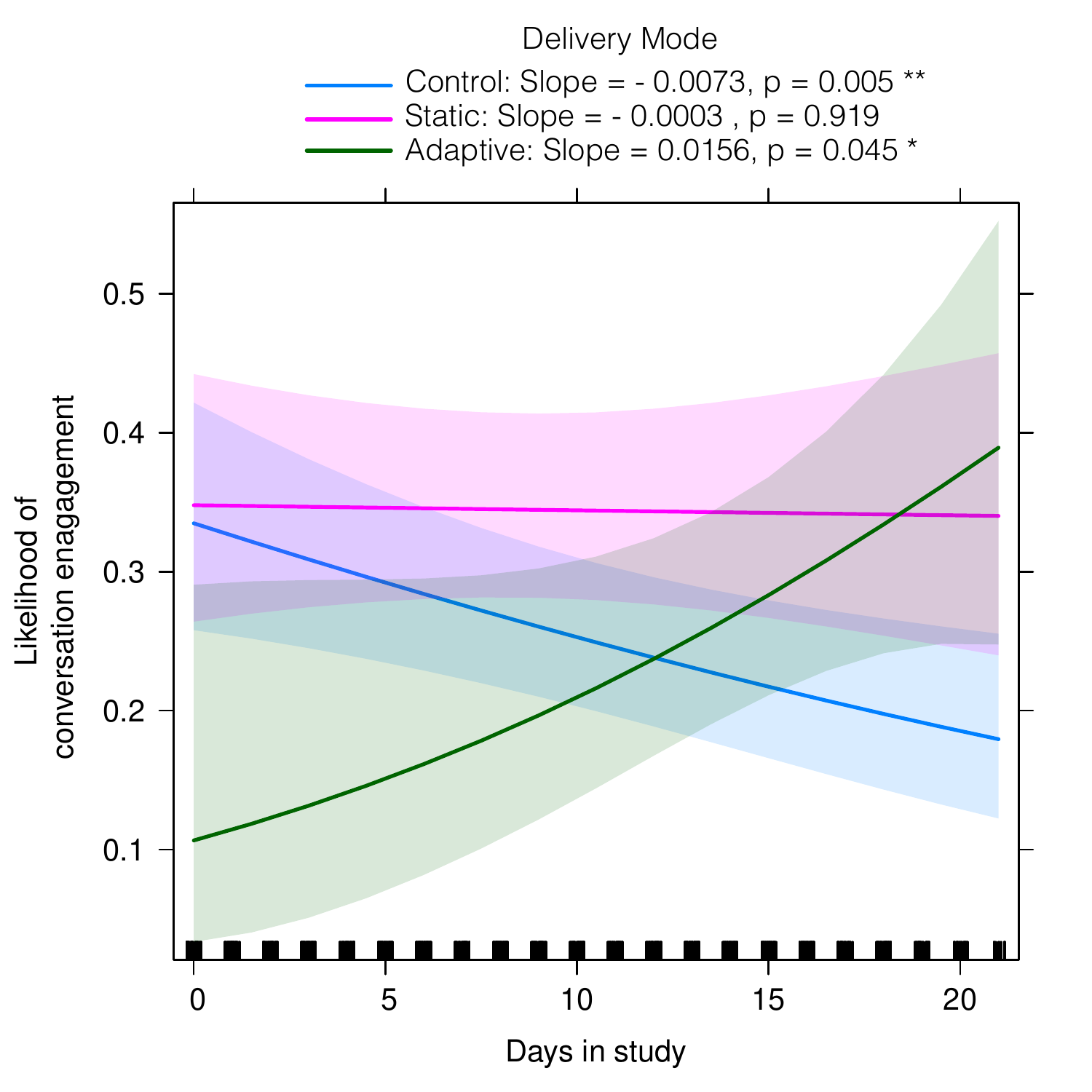} }}%
      \qquad
    \subfloat[Effect of the different models on the overall response rate over time.]{{\includegraphics[width=0.4\linewidth]{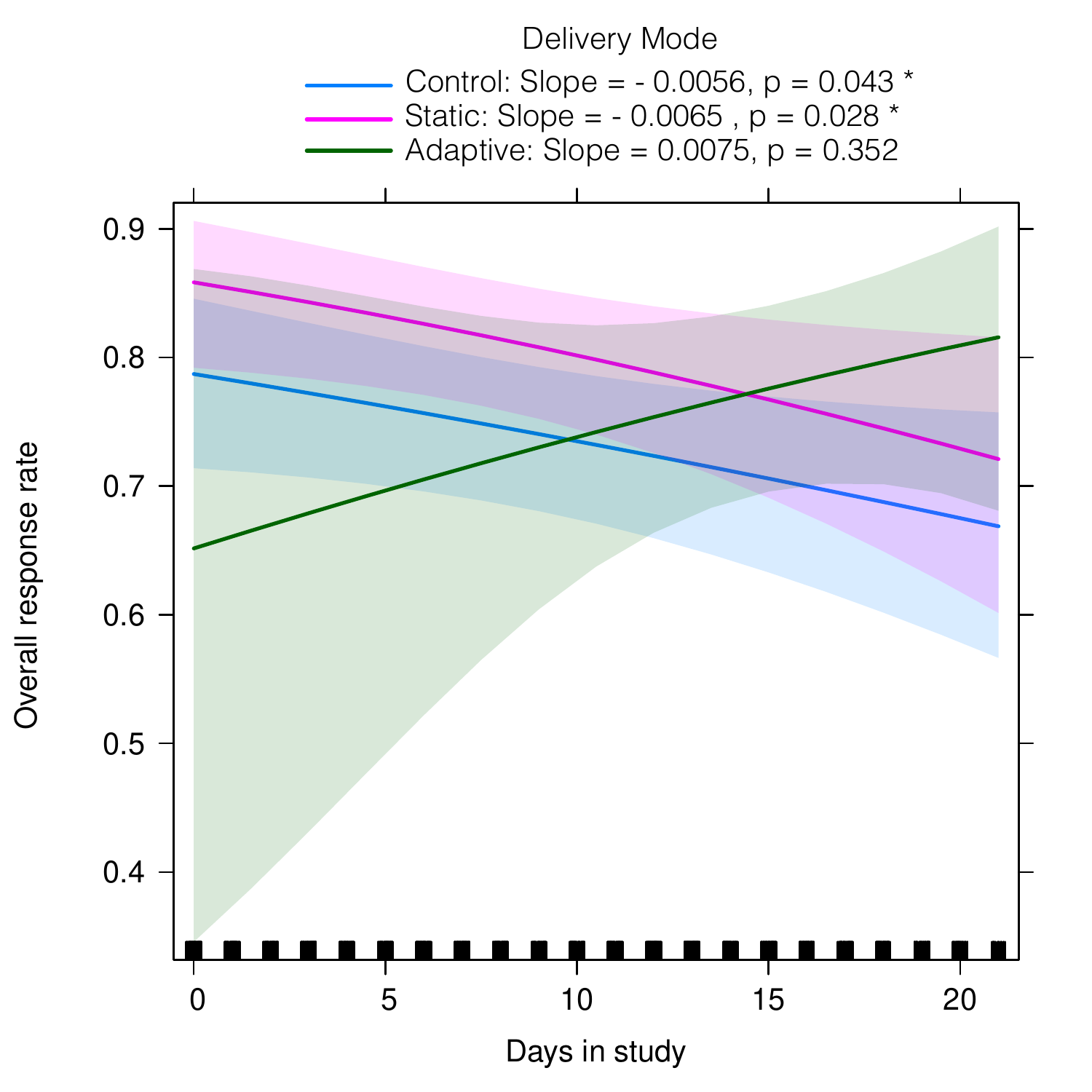} }}%
    \qquad
    \subfloat[Effect of the different models on the response delay over time.]{{\includegraphics[width=0.4\linewidth]{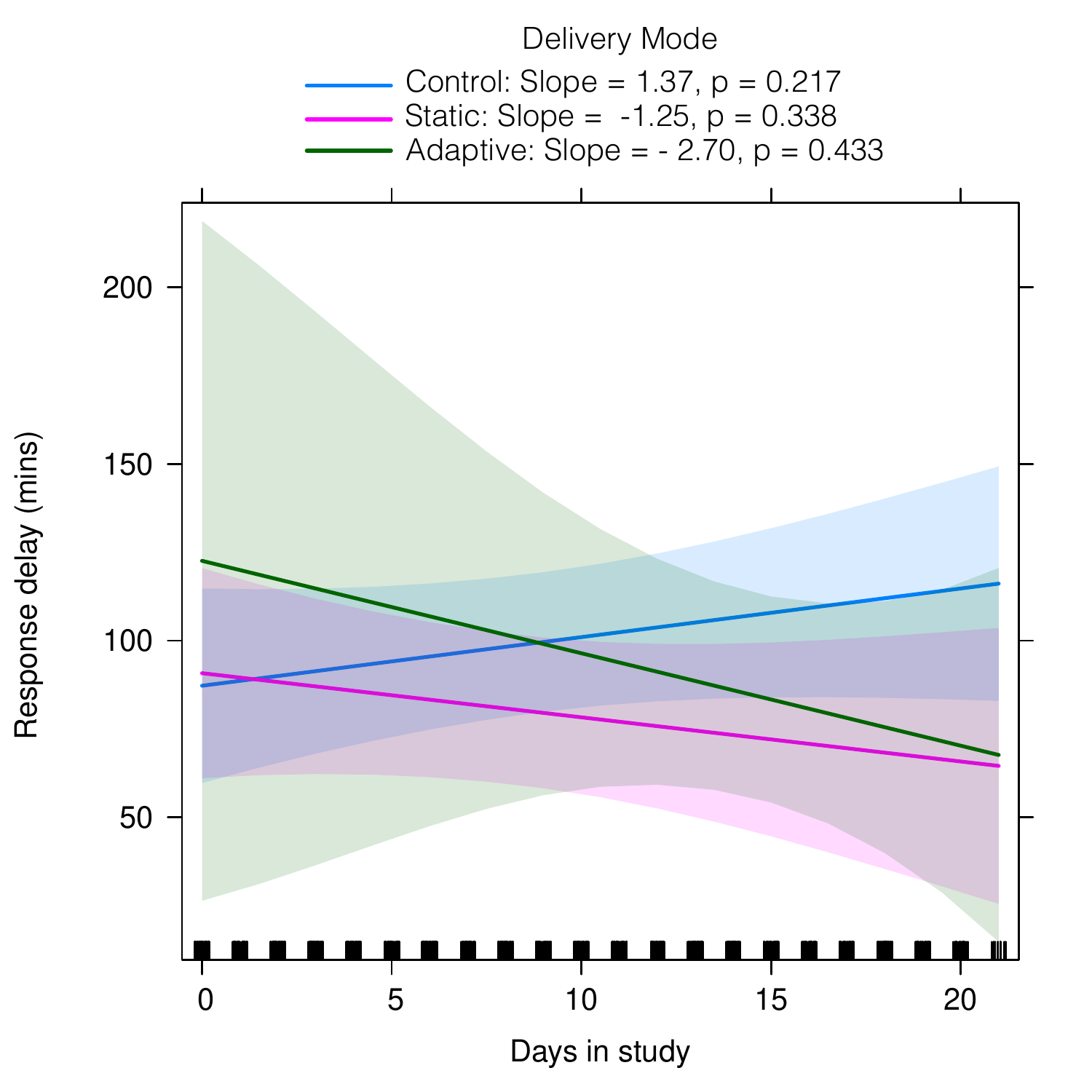} }}%

    \caption{\color{red}The performance over time of the models on the receptivity metrics. The adaptive model was only activated starting day 8; the dotted lines represent the projection of the trend for the adaptive model from day 1 to day 7.}%
    \label{fig:overtime}
\end{figure}

Looking at the conversation-engagement rate (Figure~\ref{fig:overtime}b), the control model showed a significant downward trend ($p=0.005$), suggesting that for messages delivered by the control model, the conversation-engagement rate declined over the course of the study. The static model had a slight, yet insignificant, downward trend, similar to the just-in-time response rate. The adaptive model had a significant positive trend ($p = 0.045$), with a slope of $0.0156$, which translates to a $1.56$ percentage-point increase in conversation-engagement rate each day. This result further supports our expectation that the adaptive model would be able to learn from the personalized data and improve itself. Although conversation engagement was not the primary metric of focus, it still is an important part of the overall concept of receptivity, especially to chat-based interventions like Ally.

Next we evaluated the effect on the overall response rate (Figure~\ref{fig:overtime}c). Consistent with the two previous metrics, the control model had a significant downward trend, suggesting that overall response rate for messages delivered by the control model decreased over the course of the study ($p=0.043$). The static model was more interesting: although the just-in-time response rate to messages from the static model did not reduce over the days, there was a decline in the overall response rate ($p=0.028$) . The static model led to a higher just-in-time response rate than the control model, and kept it high as the just-in-time response rate for the control model fell over time. However, just like the control model, if the participant did not respond just-in-time, the overall response rate reduced over the days, following the trend of attrition common in mobile sensing and mHealth studies~\cite{wang:studentlife14, tseng:pilot16}. The adaptive model was able to maintain a satisfying response rate throughout.%

Finally, for response delay (Figure~\ref{fig:overtime}d), we did not discover a trend that was significant in any of the three models. We make several encouraging observations, however. At the beginning, the response delay for the control and static models was approximately the same, but they diverged as the study progressed; the control model had an increasing trend, whereas the static model had a decreasing trend, suggesting that the participants' response time to the static model improved as the study progressed. The adaptive model shows an even steeper improvement, with a slope of $-2.70$, suggesting an improvement of almost 3 minutes each day. It is important to note that the results for response delay were not significant, and more research is needed.}

\subsection{Summary of Results}
Based on our analysis of RQ-1 and RQ-2, our first two hypotheses (H-1 and H-2) were partially true. The \emph{static\/} model led to a significant improvement over the control model in just-in-time response rate, overall response rate, and conversation engagement. The \emph{adaptive\/} model led to slight improvements over the control model, though the results were not significant. The results for RQ-3 provide some insights on the lack of significant improvement in the adaptive model. As shown in Figure~\ref{fig:overtime}, the receptivity to the adaptive model started similar to that of the control model, but it kept improving as the study progressed, although the average over the study was not significantly higher than the control model.

    Furthermore, we observed that after Day 19, the just-in-time response rate from the adaptive model was higher than that from the static model. Similarly, after Day 18, the conversation engagement from adaptive model was higher than that from the static models. While these are encouraging results that indicate the adaptive model continued to improve over time, in our short study we could not observe the trend over a longer period of time.  We hypothesize the adaptive model would continue to improve beyond Day~21, outperforming the static model, although it would eventually plateau. We anticipate future research will be able to test this hypothesis.

\section{Discussion} %
\label{sec:discussion}
In our work, we show that machine-learning models can be used to predict receptivity to interventions. We found that intervention alerts delivered by a \emph{static\/} model performed significantly better than delivering interventions at random times. Further, we found that receptivity to an adaptive model -- which learnt user specific features over the course of the study -- improved as the study progressed. In this section, we discuss the implications of our results in three broad categories, along with the limitations, and future directions for each.

\subsection{Domain-specific Models to Detect Receptivity}

In prior work, Mehrotra~et~al. found that \textit{notification category} was one of the top features to determine whether users would react to notifications~\cite{mehrotra2015designing}, thus highlighting the importance of \emph{who\/} or \emph{what\/} is sending an alert. Along with \emph{who\/} and \emph{what\/} is sending an alert, Visuri~et~al.\  showed that the actual content of a prompt is also important~\cite{visuri:notification19}. Using a semantic analysis of a notification's content (along with contextual features) drastically improved the detection of opportune moments to smartphone notifications~\cite{visuri:notification19}. 
These findings show the importance of the content of a notification and could potentially affect the generalizability of ML models to different interventions.

In our work, we considered receptivity to the initial greeting  message.
The initial message was a generic greeting message like ``Hello [participantName]'' or ``Good morning [participantName]''. Only after the participants replied back to the greeting did the actual intervention conversation start. Since the receptivity to interventions was not affected by the content of intervention, we argue that our results and models could be generalizable to other \ac{JITAI}s with a similar level of intervention engagement, or to other \ac{JITAI} that use chatbots for digital coaching. {\color{red}However, more deployments with different intervention types are needed to get a better sense of generalizability of our results. } 

It is important, however, to be mindful of the findings by Mehrotra~et~al.~\cite{mehrotra2015designing} and Visuri~et~al.~\cite{visuri:notification19}. We do not know whether and how our results might generalize to interventions that are more involved, or require the user to actively perform some task, e.g., to take 10 deep breaths or to take some medications. More research is needed, especially in the domain of cognitive availability, to explore how receptivity changes with intervention burden. 

\subsection{Dependence on Intervention Design and Effectiveness}
Based on the definition of \ac{JITAI} proposed by Nahum-Shani~et~al.~\cite{nahum2016just}, \emph{receptivity\/} can be considered a \emph{tailoring variable}, which indicates whether a user is available to receive an intervention at any given time. For receptivity to be an informative parameter, it is imperative that the actual intervention being delivered is effective. Intervention designs that are ineffective or cumbersome might not lead to engagement, regardless of whether a user is available to receive the information. 

We used an intervention design similar to that proposed by Kramer~et~al.~\cite{kramer2018investigating, kramer:step-goals}. Although evaluating the effectiveness of interventions is beyond the scope of this paper, Kramer~et~al. found that when compared to a baseline (no intervention) period, on average the participants in their study significantly improved their daily step counts by 438 steps~\cite{kramer:step-goals}. 

Another aspect to consider is the determination of a time-window for receptivity. Based on prior work by K\"{u}nzler~et~al.~\cite{kunzler:receptivity2019} and Mehrotra~et~al.~\cite{mehrotra2015designing}, in our exploratory work we considered receptivity to interventions if a participant responded to an intervention message within 10-minutes of delivery. We believe that the choice of the time-window to determine receptivity would eventually depend on the intervention type and what is considered as an acceptable receptive duration for that intervention. Some time-critical intervention designs might require response within a 1-minute window to be considered ``receptive'', whereas others might consider a 1-hour response time acceptable. Depending on what is considered receptive, the models and their subsequent performance could significantly change. {\color{red}As a preliminary evaluation, we used the previous Ally data and tested feasibility of models with other window sizes (1 minute, 3 minutes, and 5 minutes). We found that for each of these windows, the machine-learning models consistently out-performed the random baseline classifier. For instance, the F1-score of the random classifier for the 1-minute window was 0.11, whereas the SVM model had an F1-score of 0.18, an improvement of over 60\%. Hence, it seems that training models for different time-windows is feasible, and we plan to explore receptivity to varying time-windows further in future work and delve deeper into the influence of intervention design and intervention effectiveness on receptivity-detection models.}

\subsection{Personalized Models to Detect Receptivity}
In prior work, Morrison~et~al.\  deployed a mobile stress-management intervention system, where they built personalized Naive Bayes models to deliver interventions at \emph{opportune\/} moments~\cite{morrison2017effect}. They found, however, that there was no difference in how the participants interacted with notifications at opportune times vs. those delivered at random times. One reason could be that they used a static personalized model trained after an initial ``learning period.'' It is possible that the model did not have enough data to be adequately trained, i.e., the cold-start problem.

Our results show that the \textit{adaptive} model (which followed a dual-model approach) started with poor performance, but the performance improved as more data was available. Further, even in our results, we observed that the adaptive model did not perform significantly better than the control model if we considered the entire study duration (RQ-1 and RQ-2). It was when we evaluated the day-by-day performance (RQ3), did we notice the increasing trend. These results suggest that personalized models can improve over time with more data. Further, given that the static model performed significantly better over the control model, a dual-model solution could be appropriate to deal with the cold-start problem. We discuss some potential approaches in the next section.

\subsection{Building better Models}
As an exploratory study, our work used several \emph{in-the-moment\/} contextual features, like phone battery level, device interaction, date/time, and physical activity. Since our models led to significantly higher receptivity than a random model, it is natural to question whether incorporating more features could further improve receptivity. In our work, we did not consider the location of a user; although in our prior work we did not find any significant association between a user's location and his/her receptivity, several other works have noted the effect of location type on notification engagement~\cite{sarker2014assessing, pielot2017beyond}. Other features like demographic information~\cite{kunzler:receptivity2019, pielot2015attention, pielot2017beyond} and personality traits~\cite{mehrotra2016my, kunzler:receptivity2019} have shown to correlate with receptivity and interruptibility. 

Further, our results show that the performance of the \emph{adaptive\/} model improved as the study progressed. This is one of the first works to show the improvement of an adaptive model to detect receptivity in a real-world deployment, and highlights the potential of deploying adaptive models that can be tuned to each participant for optimal performance in the long term. In our work, the adaptive model included two models (P1 \& P2), and both had equal weights. Given the promising results, we now hope to explore models with adaptive weights based on the relative number of data points to train each model; thus, initially P1 would have a higher weight, but over time as P2 accumulates more data points its weight would increase and eventually be higher than P1. In the future, we plan to understand how adding new features would affect model performance, and explore what type of adaptive model might be best, how to best change the weights of a personalized model, and also how to determine the optimal number of days to enable the personalized component.

Finally, although our results are promising, they are still preliminary. We had 83 users in our study who participated for only 3~weeks; most behavior change programs last longer than 3~weeks. Hence, more research is needed to evaluate model performance and how receptivity changes over a longer period of time. 

{\color{red}
\section{Implications on JITAI Design}
As highlighted in Section 3.2, a JITAI has six key elements: a distal outcome, proximal outcomes, decision points, intervention options, tailoring variables, and decision rules. Our results show that it is indeed possible to detect receptivity. Hence future studies could design JITAIs such that the intervention components and decisions rules can account for receptivity as a tailoring variable before making a determination on whether to deliver an intervention.

In our work, we considered receptivity as a binary outcome, i.e., a person is either receptive or not. This is just the first step towards enabling effective delivery of interventions. We argue that \textit{receptivity} is a spectrum and not an absolute yes/no. It could be possible that -- in a given moment -- a person is receptive to a particular type of intervention and be non-receptive to a different type of intervention. Given the promising results in our study, we lay solid groundwork for future researchers to move forward to other dimensions of receptivity. The treatment of receptivity as a spectrum would enable intervention designers to decide not only \emph{if} an intervention should be delivered but also \emph{what} interventions to deliver in that moment -- JITAIs could be developed that consider the degree of vulnerability (tailoring variable), the level of receptivity (tailoring variable), and the expected effectiveness of various interventions (intervention options) and decide which intervention to maximize the distal and proximal outcomes.
}

\section{Summary and Conclusion} %
\label{sec:conclusion}
We conducted a study in which we deployed machine-learning models that detect momentary receptivity in the natural environment. 
We leveraged prior work in receptivity to \ac{JITAI}s and deployed a chatbot-based digital coach~-- \appname~-- that provided physical-activity interventions and motivated participants to achieve their step goals. The \appname~app was available for iOS and included two machine-learning models that used information about the app user's context to predict whether that person was likely to be receptive at that moment. We used two types of machine-learning models: (a)~a static model, that was built before the study started and remained constant for all participants and all time; and (b)~an adaptive model, that continuously learnt the receptivity of individual participants and updated itself as the study progressed. For comparison, we deployed (c)~a control model that sent intervention messages at random times. The choice of model to be used for delivery was randomized for each intervention message. We observed that messages delivered using the machine-learning models led up to a 40\% improvement in receptivity as compared to the control model. Further, we evaluated the temporal dynamics of the different models over time, and observed that while the receptivity to messages from the control model declined, receptivity to messages from the adaptive model increased over the course of the study.
 \ifanonymized  %
\else
\section*{ACKNOWLEDGMENTS}

The authors thank the anonymous reviewers and editors for their helpful feedback. This research results from a joint research effort of the Centre for Digital Health Interventions at ETH Z{\"u}rich and the Center for Technology and Behavioral Health at Dartmouth College.  It was supported by the NIH National Institute of Drug Abuse under award number NIH/NIDA P30DA029926, and by CSS Health Insurance, Switzerland. The views and conclusions contained in this document are those of the authors and should not be interpreted as necessarily representing the official policies, either expressed or implied, of the sponsors.

 \fi

\ifacm    
    \bibliographystyle{ACM-Reference-Format}		%
\else
    \bibliographystyle{plain}
\fi
\bibliography{main} 			%
\end{document}